\documentclass[a4paper, amsfonts, amssymb, amsmath, reprint, showkeys, nofootinbib, twoside,superscriptaddress,prx,citeautoscript,longbibliography]{revtex4-2}
\usepackage[english]{babel}
\usepackage[utf8]{inputenc}

\usepackage[colorinlistoftodos, color=green!40, prependcaption]{todonotes}
\usepackage{amsthm}
\usepackage{mathtools}
\usepackage{physics}
\usepackage{xcolor}
\usepackage{graphicx}
\usepackage[left=16mm,right=16mm,top=20mm,bottom=22mm,columnsep=15pt]{geometry} 
\usepackage{adjustbox}
\usepackage{placeins}
\usepackage[T1]{fontenc}
\usepackage{lipsum}
\usepackage{csquotes}

\usepackage[pdftex, pdftitle={Article}, pdfauthor={Author}]{hyperref} 
\usepackage{float}

\def\kbt{k_{\mathrm{B}}T}

\newcommand{\comment}[1]{}

\begin{document}
\title{Reactive path ensembles within nonequilibrium steady-states}
\author{Aditya N. Singh}
\affiliation{Department of Chemistry, University of California, Berkeley, CA 94720, USA}
\affiliation{Chemical Sciences Division, Lawrence Berkeley National Laboratory, Berkeley, CA 94720, USA}

\author{David T. Limmer}
\email[Electronic mail: ]{dlimmer@berkeley.edu}
\affiliation{Department of Chemistry, University of California, Berkeley, CA 94720, USA}
\affiliation{Chemical Sciences Division, Lawrence Berkeley National Laboratory, Berkeley, CA 94720, USA}
\affiliation{Materials Science Division, Lawrence Berkeley National Laboratory, Berkeley, CA 94720, USA}
\affiliation{Kavli Energy Nanoscience Institute at Berkeley, Berkeley, CA 94720, USA}

\date{\today} 

\begin{abstract}
The modern theory of rare events is grounded in near equilibrium ideas, however many systems of modern interest are sufficiently far from equilibrium that traditional approaches do not apply.  Using the recently developed variational path sampling methodology, we study systems evolving within nonequilibrium steady states to elucidate how reactive processes are altered away from equilibrium. Variational path sampling provides access to ensembles of reactive events, and a means of quantifying the relative importance of each dynamical degree of freedom in such processes. With it, we have studied the conformational change of a solute in an active bath. We illustrate how energy injection generically enhances the rates of rare events, even when energy is not directed into specific reactive modes. By studying the folding and unfolding transitions of a grafted polymer under shear, we illustrate how nonequilibrium reactive processes do not follow gradient paths due to the emergence of persistent currents. The breaking of detailed balance allows for the mechanisms of forward and backward reactions to be distinct, enabling novel pathways to be explored and designed, and states unstable in equilibrium to become stabilized kinetically away from it. The analysis presented in this work establishes some basic principles for nonequilibrium reactive events, and is made possible by the use of a numerical method that does not invoke proximity to equilibrium or requires strong prior assumptions about the mechanism of reaction.
\end{abstract}
\maketitle
\section{Introduction}

Systems evolving nonequilibrium steady states (NESS) have their thermodynamics entangled with their kinetics.  As a consequence, these systems can exhibit exotic behavior as they are not constrained by equilibrium conditions such as detailed balance.~\cite{vicsek1995novel, omar2021phase, cates2015motility, palacci2013living, tan2022odd, marchetti2013hydrodynamics,kinosita2000rotary,lau2007nonequilibrium,zandi2003drives,sharma2018non,alexander2006shear,schneider2007shear}  While much attention has been paid to emergent structures and local dynamics within NESS, comparably less is known regarding the dynamics accompanying transitions between metastable nonequilibrium states.
Transitions between metastable states in NESS have been observed to exhibit accelerated kinetics compared to their equilibrium counterparts, with minimal perturbation of their steady-state density~\cite{mallory2018active,cho2023tuning,walther2008janus,kuznets2021dissipation,das2022nonequilibrium,osat2024escaping}. Furthermore, metastability within NESS can emerge through mechanisms that are dynamic in nature and cannot necessarily be rationalized through energetic or entropic considerations that are used to describe these phenomena in thermal equilibrium ~\cite{tan2022odd, batton2024microscopic,alexander2006shear,redner2016classical,omar2023mechanical}. Although such phenomenology is ubiquitous in the physical and biological sciences, few guiding principles exist to explain or predict such behavior. 

In this work, we study systems driven into NESS in order to understand how metastablity differs between systems in and out of equilibrium.  We employ ideas from kinetic theory and stochastic optimal control to understand why the addition of active forces that are dissipative and nonspecific tend to accelerate kinetics, and how metastability can emerge from unbalanced probability currents. The  insights in this work are largely made possible due to the recent developments of variational path sampling~\cite{singh2023variational,singh2024splitting}, which enables a dynamically consistent coarse-graining of NESS, where the importance of persistent fluxes and breakdown of detailed balance are actively taken into consideration. 

The study of interacting metastable systems involves the investigation of rare reactive events and requires computational methods for importance sampling and of  coarse-graining of high dimensional systems. Through development of early kinetic theories based on configurational averages ~\cite{chandler1978statistical,hanggi1990reaction,kramers1940brownian,eyring1935activated}, coupled with the availability of a wide variety of configurational importance sampling ~\cite{van2003novel,bolhuis2002transition,barducci2011metadynamics,torrie1977nonphysical} and coarse graining methods~\cite{vanden2010transition,metzner2009transition,thiede2019galerkin,coifman2008diffusion,strahan2021long,prinz2011markov}, it has become tractable to efficiently characterize reactive phenomena in complex systems that obey detailed balance. 
For systems that are driven far from equilibrium due to the constant injection of energy, the toolkit is rather sparse. Due to the breakdown of detailed balance, rate theories involve path averages~\cite{freidlin1998random,zakine2022minimum,heller2024evaluation}, and due to the lack of a known steady state distribution, the availability of importance sampling methods is restricted~\cite{warmflash2007umbrella,allen2009forward,das2022direct,das2019variational}. The breakdown of detailed balance and accompanying finite rate of entropy production allows for the dominant reactive pathway from one state to the other to differ from the path taken the other way. This has made it challenging to understand how perturbations from nonequilibrium forces affect rates of rare events through arguments based on a configurational distribution. Methods available for coarse-graining systems in NESS are also limited~\cite{knoch2017nonequilibrium,helfmann2020extending,strahan2023inexact,strahan2023predicting}, and consequentially, standard approaches employ equilibrium or near-equilibrium theories to investigate reactive phenomena in systems far from equilibrium. These approaches can be accurate in limiting cases, however they neglect the importance of steady probability currents that underlie the properties of systems far from equilibrium.~\cite{fang2019nonequilibrium,fodor2016far,das2022nonequilibrium} 

%The dynamical nature of the steady-state implies that steady currents can lead to changes within the steady-state distribution and stabilize or destabilize specific states in ways that cannot be easily rationalized through equilibrium arguments of entropy or energy.  

Here we explore two problems within the intersection of kinetics and nonequilibrium steady states to illustrate how probability currents and the breakdown of detailed balance can govern metastability and rates of rare events. First, we consider the effect of adding active forces to a passive reaction using a model of active Brownian motion. We find across a range of systems that activity tends to enhance kinetics and rates of rare events. To understand the nature of this enhancement, we develop a theory using simple tools from stochastic optimal control, spectral theory and transition path theory. By validating our theory across a range of systems with varying complexity, we find that the nature of the enhancement is closely related to the breakdown of detailed balance. Specifically, we find that the addition of activity gives rise to multiple enhanced and suppressed pathways, and due to the breakdown of detailed balance, the systems can selectively transition through the enhanced pathways and effectively enhance the rate of reaction. 

Second, we consider the effect of adding non-conservative shear forces to systems that does not exhibit metastability. We find across a range of systems that shear flow gives rise to steady probability currents that can be leveraged to engineer emergent metastable states. Shear induced stability can be understood through the interplay between steady currents and barriers that trap them in order to accumulate probability.  The generality of this phenomenon is illustrated through a variety of models of increasing complexity. The systems presented in this work exhibit a variety of exotic and novel phenomena which we are able to decipher effectively through the application of our recently developed method~\cite{singh2023variational,singh2024splitting} with minimal \textit{a-priori} assumptions of the systems and reactive events.

\section{Dynamical reweighting and control}
Most of the results and analysis we will perform leverage recent results on stochastic optimal control and dynamical reweighting~\cite{kuznets2021dissipation,das2022direct}. 
Stochastic optimal control has offered a particularly promising means of addressing questions of nonequilibrium systems, as it provides mechanical interpretations to rare nonequilibrium dynamics~\cite{kuznets2021dissipation,das2022direct}. The results presented here are largely derived from previous results, so we will only summarize some of the key definitions and relations. More information on the derivations of these results can be found in Refs.~\onlinecite{kuznets2021dissipation, das2022direct,singh2023variational,singh2024splitting}. For simplicity we will consider a system with $N$ degrees of freedom denoted $\mathbf{r}$ evolving from an overdamped Langevin equation equation of motion in $d$ dimensions, 
\begin{equation}\label{eq1}
    \gamma_i \dot{r}_i = F_i(\mathbf{r}) + \eta_i 
\end{equation}
where $\dot{r}_i$ is the time derivative of the position of the $i$'th degree of freedom, $F_i(\mathbf{r})$ is the drift along that coordinate that need not be given by the gradient of a potential. The force, $\eta_i$ denotes delta correlated Gaussian noise with variance $\langle \eta_i(t) \eta_j(t') \rangle = 2\gamma_i \kbt \delta (t'-t) \delta_{ij} $ where $\langle \hdots \rangle$ denotes a thermal average, and $\kbt$ denotes  Boltzmann's constant times the temperature. 

Our primary interest is to consider systems that exhibit metastability, which is a statement of a gap within the spectrum of the Fokker-Planck operator that describes the evolution of the probability density generated by Eq. \ref{eq1}.~\cite{bovier2016metastability} Specifically, we focus on bistable systems, an assumption applicable to a range of systems. For such systems, the primary quantity of interest is the rate of reaction $k$  between the two stable regions denoted $A$ and $B$ that can be related to the conditioned transition probability over some finite time $t_f$ under an assumption of a separation of timescales~\cite{chandler1978statistical} 
\begin{equation}
    k t_f \approx \frac{\langle h_A(0) h_B(t_f)\rangle}{\langle h_A \rangle} = \langle h_B(t_f) \rangle_A
\end{equation}
where $h_A$ and $h_B$ denote indicator functions identifying the $A$ and $B$ region, 
\begin{equation}
     h_X[\mathbf{r}(t)] = \begin{cases} 1 \qquad \mathbf{r}(t)\in X \\ 0 \qquad \mathbf{r}(t)\notin X   \end{cases}
\end{equation}
where $X=\{A,B\}$. The existence of a gap in the spectrum of the Fokker-Planck operator ensures that this separation of timescales exists. We assume that $A$ and $B$ are non-intersecting sets of configurations. 

Metastability implies that trajectories, $\mathbf{X}(t_f)=\{\mathbf{r}(t)\}_{0 \leq t \leq t_f}$, of duration shorter then $1/k$ are unlikely to exhibit a reactive event between the sets of configurations $A$ and $B$. As in equilibrium ensemble reweighting theory, one can ask how an observable, like $k$, changes with parameters of the system. In this case, we can understand the effect of perturbing the original system with an additional force $\mathbf{\lambda}(\mathbf{r},t)$, that enters the equation of motion
\begin{equation}\label{eq1b}
    \gamma_i \dot{r}_i = F_i(\mathbf{r}) + \lambda_i(\mathbf{r}) + \eta_i 
\end{equation}
additively to the original force. Through Girsanov's theorem~\cite{kuznets2021dissipation}, we can relate the rate of the perturbed system $k_\lambda$ to that of the original system
\begin{equation}\label{eq:rate_OM}
    \ln k_\lambda / k_0 = \langle e^{-\Delta U_{\lambda}} \rangle_{B|A, 0} = \langle e^{-\Delta U_{\lambda}} \rangle_{B|A, \lambda}^{-1}
\end{equation}
where the relationship is given by the difference stochastic action's $\Delta U$ averaged within the reactive trajectory ensemble of the original and perturbed dynamics denoted by $\langle \rangle_{B|A,0}$ and $\langle \rangle_{B|A,\lambda}$ respectively. While such a statement is general for any stochastic dynamics for which the dynamics with and without the applied force span the same space of possible trajectories, for the overdamped Langevin dyanmics, the change in action has a simple form
\begin{eqnarray}\label{eq:OM}
    \Delta U_{\lambda} &=& \ln \frac{P_\lambda[\mathbf{X}(t_f)]}{P_0[\mathbf{X}(t_f)]} \\
    &=& -\sum_i^N \int_0^{t_f} dt \frac{\lambda_i^2 - 2\lambda_i (\gamma_i \dot r_i-F_i)}{4\gamma_i \kbt} \nonumber
\end{eqnarray}
where $P_0[\mathbf{X}(t_f)]$ and $P_\lambda[\mathbf{X}(t_f)]$ are the probabilities of observing a trajectory of length $t_f$ in the original and the perturbed systems, respectively.

The reweighting relation in Eq. ~\ref{eq:rate_OM} is general for any applied force, $\boldsymbol{\lambda}=\{\lambda_1, \dots, \lambda_{dN}\}$. Through the convexity of the exponential, a variational bound can be deduced from Jensen's inequality,
\begin{equation}\label{eq:rate_bound}
   %\langle \Delta \Gamma_{\lambda}  \rangle_{B|A, 0} \leq 
   \ln k_\lambda / k_0 \leq   \langle \Delta U_{\lambda} \rangle_{B|A, \lambda}
   \end{equation}
 where the log-ratio of rates with and without the applied force is bounded by averages of the change in action in the reactive ensemble under the applied force. An equivalent bound holds for the ensemble with the applied force. This bound is saturated in two limits. First it follows from a cumulant expansion of Eq.~\ref{eq:rate_OM}, that the bound is saturated within a linear response region for the rate. Linear response is expected to be valid for forcing that are small relative to the maximum force encountered in the reaction and independent of the details of the extra force~\cite{kuznets2021dissipation}. The other limit for which the bound is saturated is for a specific force, at arbitrary strengths.  The existence of an optimal force that saturates the bounds in Eq.~\ref{eq:rate_bound}, denoted $\boldsymbol{\lambda}^*$, is given by the backward Kolmogorov equation. It is a time-dependent conservative force, often termed as the optimal controller of the bridge process~\cite{das2022direct,majumdar2015effective,chetrite2015nonequilibrium,chetrite2015variational}.  It is a force that ensures reactivity within finite time $t_f$, such that $k_\lambda = 1/t_f$, and 
\begin{equation}
 \ln  k_0 =- \sum_{i}  \int_0^{t_f} dt \frac{ \langle \lambda_i^2 \rangle_{B|A,\lambda^*}}{4 \gamma_i \kbt } 
 \end{equation}
the rate in the original system is given by an average over the optimal force. This provides a simple mechanical interpretation of the rate of a rare event as the smallest added force to ensure reactivity. The linearity of the expression, where the rate is written as a sum over contributions to each degree of freedom, permits a natural decomposition of the rate into individual degrees of freedom that have to be activated or forced in order for the rare fluctuation to occur. 

While in general the contributions to the rate are delocalized over many degrees of freedom, we can perform a coordinate transformation to try to localize those contributions to a few specific modes. Transforming the force through $ \tilde{\boldsymbol{\lambda}} = \mathbf{J}^{-1} \cdot {\boldsymbol{\lambda}}$ where $\mathbf{J}$ is the Jacobian of the transformation, we can  express the rate as
\begin{equation}\label{eq:decomp}
 \ln  k_0 =- \sum_{i} \langle  \Delta \bar U_{\lambda^*}^{i}  \rangle_{B|A,\lambda^*}
 \end{equation}
   with 
\begin{equation}
    \Delta \bar U_{\lambda^*}^{i} = \sum_{j} \int_0^{t_f} dt \frac{\tilde \lambda_i \Gamma^{-1}_{ij} \tilde \lambda_j}{4 \kbt} 
\end{equation}
where $\Gamma^{-1}_{ij}= \sum_k J_{ki} J_{kj}/\gamma_k$ is a friction weighted metric tensor~\cite{singh2023variational}. This transformation allows us to express the rate as a sum of contributions from collective coordinates, whose transformation matrix can be optimized to uncover mechanisms of rare events, as those collective coordinates that need to be forced to ensure reactivity.  In the following, we will use the relationship between rates and mechanical forces and the decomposition of the rate to clarify how nonequilibrium forces enhance rates and alter stability.

\section{Rate enhancement}

It is routinely observed that the introduction of nonequilibrium forces can speed up kinetics. This is most obvious when a control force is added to the system to directly facilitate a transition, as in single molecule pulling experiments ~\cite{bustamante2004mechanical}. However, it has been observed that transverse forces that leave the steady-state invariant can increase mixing times of Markov chains.~\cite{kolchinsky2024thermodynamic} It has also been demonstrated that the application of nonreciprocal interactions can increase the kinetics of nucleation~\cite{cho2023tuning}, and facilitate escape from kinetic traps to aid self assembly~\cite{das2022nonequilibrium,osat2024escaping,batton2024microscopic}. Nonequilibrium speed limits and thermodynamic uncertainty relationships have offered some guidance to understand how dissipation can alter kinetic processes~\cite{gingrich2017fundamental}, and the recently discovered heat bound on rate enhancement suggests that coupling to non-directed dissipative processes is expected to universally increase rates of rare events~\cite{kuznets2021dissipation}. Eqs. \ref{eq:rate_OM} and $\ref{eq:rate_bound}$ offer an exact form and a strong bound for the rate enhancement of a system perturbed by an external force respectively. However, since these relations require averaging of stochastic variables within a reactive trajectory ensemble, further insight is often intractable \emph{a-priori}, without assuming specific forms of the original system and external force.

One way to gain insight is to consider the limiting case where the inequality in Eq. \ref{eq:rate_bound} becomes an equality, that is possible when the external force is given by the optimal controller for the bridge process $\boldsymbol{\lambda^*}$. The application of this force makes all trajectories reactive with a probability of 1, but more importantly, the trajectories obtained from driving the system with this force exhibit the exact statistics of the original reactive trajectory ensemble~\cite{das2022direct,singh2024splitting}.
While the force is strongly time-dependent and singular~\cite{majumdar2015effective}, the time and spatial dependence can be decoupled within the separation of timescales~\cite{singh2024splitting}. Of relevance to this work is the result that the direction of the force is given by the gradient of a function called the committor $\grad \bar q_B(\mathbf{r})$ or the splitting probability, that has a long history within chemical physics~\cite{vanden2006towards,bolhuis2002transition}. Often termed as the ideal reaction coordinate, the committor is understood as a conditional probability to commit to the product state rather the reactant given some position in the state space. Formally, it is a function of the whole state space of the system, and solves the Backward Kolmogorov Equation~\cite{vanden2006towards}:
\begin{equation}\label{eq:bke}
    \sum_i^N \frac{F_i(\mathbf{r})}{\gamma_i} \frac{\partial}{\partial r_i} \bar q_B(\mathbf{r}) + \frac{\kbt}{\gamma_i} \frac{\partial^2}{\partial r_i^2} \bar q_B(\mathbf{r}) = 0
\end{equation}
with boundary conditions $\bar q_B(\mathbf{r}\in A) = 0$ and $\bar q_B(\mathbf{r}\in B) = 1$. The committor is computable through a range of methods ~\cite{rotskoff2022active,hasyim2022supervised,khoo2019solving,strahan2023predicting,strahan2023inexact,liang2023probing,singh2023variational}. 

The relationship between the optimal controller for the bridge problem $\lambda^*(\mathbf{r},t)$ and the committor under the separation of timescales is given by~\cite{singh2024splitting}
\begin{align}\label{eq:doob}
    \boldsymbol{\lambda}^*(\mathbf{r},t) &\approx 2\kbt \nabla_\mathbf{r} \ln ( \bar q(\mathbf{r}) + k(t_f-t)) \notag \\
    & = \phi_B(\mathbf{r},t) \mathbf{\hat q_B} (\mathbf{r})
\end{align}
where $\phi_B(\mathbf{r},t) = 2\kbt |\grad_{\mathbf{r}} \bar q_B(\mathbf{r})| / (q_{B}(\mathbf{r}) + k(t_f-t))$ is a scalar amplitude, and $\mathbf{\hat q_B} (\mathbf{r}) = \grad_{\mathbf{r}} \bar q_B(\mathbf{r})/|\grad_\mathbf{r} \bar q_B(\mathbf{r})|$ is  a unit vector in the direction of the gradient of the committor, or what we refer to as the \textit{reaction field}. The form of this vector field bears a strong similarity to the probability current of reactive trajectories defined within transition path theory ~\cite{vanden2006towards}. The motivation for writing the optimal controller in this way was specifically to reinterpret the reaction coordinate not as a single function, but as a vector field, and this interpretation admits some interesting properties. The optimal controller for the reverse process from $B\rightarrow A$, lies exactly opposite although with a different magnitude to the optimal controller for the $A\rightarrow B$. In other words, the vector field for the $B\rightarrow A$ reaction $\mathbf{\hat q_A} (\mathbf{r}) = -\mathbf{\hat q_B} (\mathbf{r})$.
Second,  any forcing that acts perpendicular to the reaction coordinate leaves the reaction unchanged in the limit of small applied forces. Within this context, the way the kinetics of the system respond to an additional force, at least perturbatively, is determined by their alignment with this field.

\begin{figure}[t]
  \centering
    \includegraphics[width=0.42\textwidth]{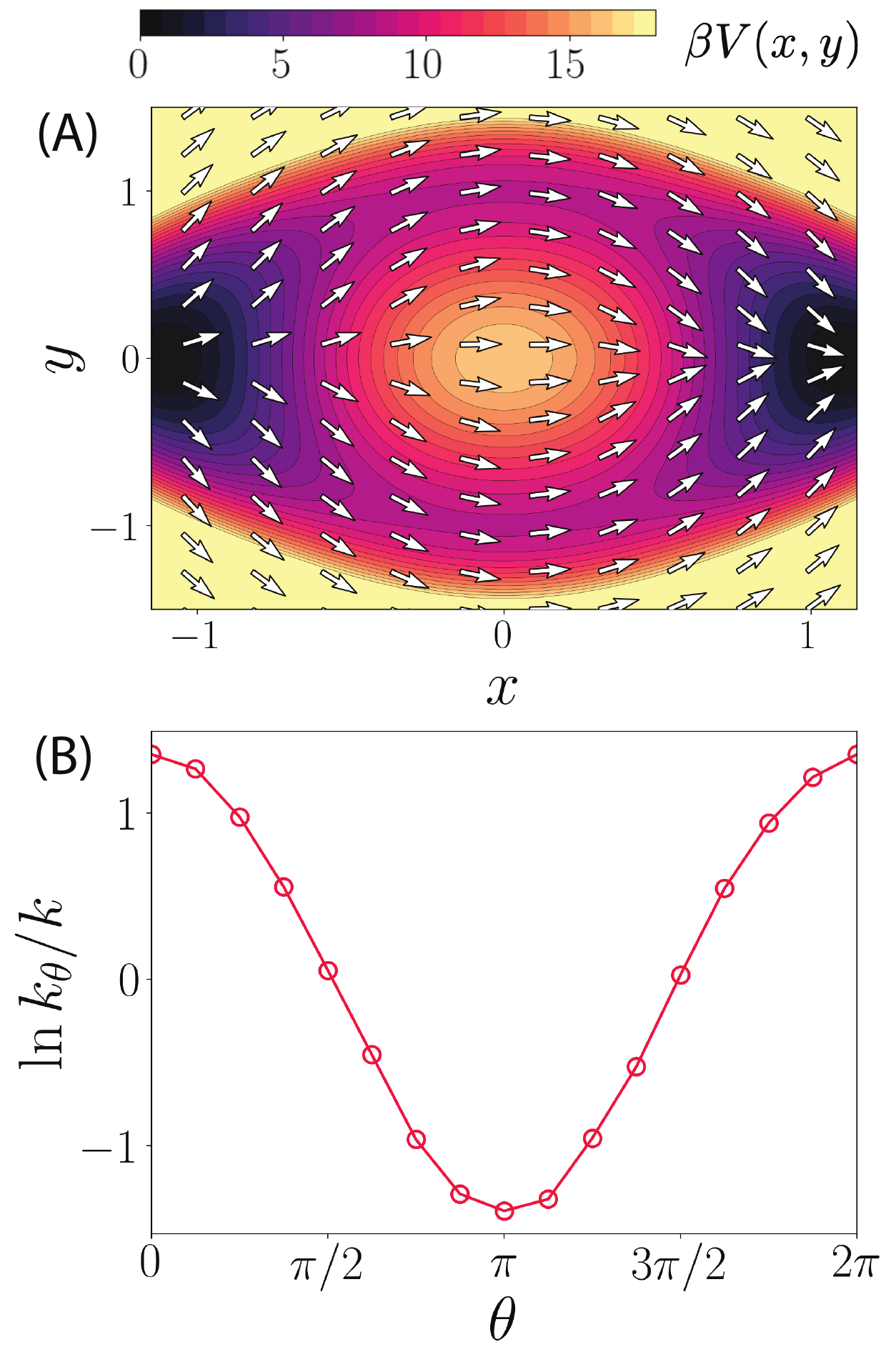}
    \caption{Effect of coupling external forces to the reaction field. (A) A contour map of the potential energy surface of a bistable system with arrows denoting the unit vector field $\hat q_B(\mathbf{r})$ corresponding to the gradient of the committor for the transition between the left and right wells denoted $A$ and $B$ respectively.  (B) Log ratio of the transition rates between the left and right wells perturbed by a linear force acting along the vector field that is rotated by $\theta$ with respect to $\hat q_B(\mathbf{r})$. $\theta=0$ and $\theta=\pi$ correspond to biasing along and exactly opposite to the reaction coordinate, respectively. }
    \label{fig:1}
\end{figure}

\subsection{Reaction coordinates and control}
To ground these ideas, we consider a Brownian particle in a bistable external potential $V(x,y)$,
\begin{equation}\label{eq:pes}
    V(x,y)/\kbt = \frac{8}{3}[6+4x^4 - 6y^2 + 3y^4 + 10x^2(y^2-1)]
\end{equation}
visulatized in Fig. \ref{fig:1} (A). The meta-stable regions $A$ and $B$ corresponding to the left and right wells are defined through the indicator functions
\begin{equation}
    h_A(\mathbf{r}) = \Theta(-x+0.85), \qquad h_B(\mathbf{r}) = \Theta(x-0.85)
\end{equation}
where $\Theta$ is the Heaviside step function. For system parameters, we choose reduced temperature $\kbt=1$, friction coefficient $\gamma=1$ and measure $x$ and $y$ in units of a lengthscale $l=1$. 

Plotted in Fig. \ref{fig:1} (A) is the unit vector function $\mathbf{\hat q_B} (\mathbf{r})$ computable from variational optimization~\cite{singh2023variational} with details discussed in Appendix \ref{appendA}.  The vector field can be understood by noting that the reaction coordinate follows the minimum action path~\cite{koehl2022sampling}, which in equilibrium is given by the minimum energy path, a gradient path following the contours of the potential. For this specific system, this path follows through the two degenerate channels located near $y=\pm1$. The direction of the gradients  closely follow the progression of the polar angle $\phi = \tan^{-1}(|y|/x)$, which was found to be an excellent descriptor of the committor for this potential,~\cite{singh2023variational} such that $q_B \approx 1/2+\erf(a \phi)$ for $a>0$ is related to the barrier height in the free energy along $\phi$.

To explore the connection between the committor and rate enhancement, we add an external force to the system with magnitude $\nu$ and direction that is rotated $\theta$ away from the commitor, $\mathbf{\hat q_B}(x,y)$, 
\begin{equation}
    \boldsymbol{\lambda} (x,y; \theta) = \nu  \mathbf{\hat e} (x,y; \theta)
\end{equation}
where $\mathbf{\hat e} (x,y; \theta) $ is defined as
\begin{equation}
    \mathbf{\hat  e} (x,y; \theta) = \left(\begin{matrix}
        \cos(\theta) & \sin(\theta) \\ 
        -\sin(\theta) & \cos(\theta) \\ 
    \end{matrix} \right)  \cdot \mathbf{\hat q_B}(x,y)
\end{equation}
such that when $\theta=0  $ or $\theta=-\pi  $, the force is parallel or anti-parallel to the reaction field respectively, and $\theta=\pi/2  $ and $\theta=3\pi/2  $ correspond to forcing perpendicular to the reaction coordinate. With this form of the external force, we choose $\nu l/\kbt = 1$ and a range of values for $\theta= [0, 2\pi]$. For each value, we estimate the rate by computing the mean first passage times from 6000 trajectories, initiated from the steady-state density of the  system with $\nu=0$ restricted to  the $A$ well. This is done by using an Euler-Murayama integrator with discretization $dt = 0.004\tau$ where $\tau$ is the reduced units for time, given by $\tau =  \gamma l^2 / \kbt$.  

The log ratio of the rate of the driven system to that of the undriven system as a function of the relative rotation $\theta$ of the external force to the reaction field $\mathbf{\hat q_B}$ is shown in Fig. \ref{fig:1} (B). The log ratio of the rates take a sinusoidal form, with maximum enhancement of $1.35 \pm 0.02$ observed when the external force lies parallel to the reaction field $\theta=0$, and maximum suppression of $1.39 \pm 0.02$ when the force acts antiparallel to the reaction field $\theta=\pi$. For the cases of $\theta=\pi/2$ and $\theta=3\pi/2$ that correspond to biasing perpendicular to the reaction field, the log ratio is found to be close to zero. Additionally, the form of the log rate ratio can be fit almost perfectly with the form $\ln k_\theta/ k = a \cos \theta$ with $a\approx1.38$. These results provide a numerical illustration of the fact that the way kinetics of the systems respond to any additional forcing is determined at least perturbatively by their alignment with a vector field related to the optimal controller. 

\subsection{Active forces and rate enhancement}
From Eqs.~\ref{eq:rate_bound} and \ref{eq:doob} and the results in the previous section, it is clear that an appropriately chosen deterministic force can enhance the rate of a reaction and that the maximum enhancement will be when the force aligns with committor. However, within linear response it is possible that any applied force, even a stochastic one, can enhance the rate. Indeed it has been observed that the addition of activity to a passive system can increase the rate of escape from metastable states at both the single particle and collective level~\cite{woillez2019activated,kuznets2021dissipation, das2022direct, caprini2021correlated}.

To explore this possibility, we replace the passive Brownian particle with an active one where the equations of motion are given by
\begin{align}\label{eq:2D_eqm}
    \gamma \dot x &= -\partial_x V(x,y) + \nu \cos(\theta) + \eta_x\notag\\
    \gamma \dot y &= -\partial_y V(x,y) + \nu \sin(\theta) + \eta_y\notag\\
    \dot \theta &= \sqrt{2D_r}\xi
\end{align}
and $\theta$ now denotes an additional degree of freedom corresponding to the director that is undergoing free diffusion with constant $D_r$ while $\xi$ is delta correlated, white noise with unit variance. This form of activity has often been used as a standard prescription to describe the equations of motion of patchy colloids~\cite{walther2008janus} under the effect of phoretic forces exerted by chemical reactions. For finite values of $\nu$, the external force cannot be written down as the gradient of the potential, and as a result, the system driven by this force will enter a nonequilibrium steady state where the stationary distribution is not necessarily a Gibbs-Boltzmann distribution. 

\begin{figure}[t]
  \centering
    \includegraphics[width=0.49\textwidth]{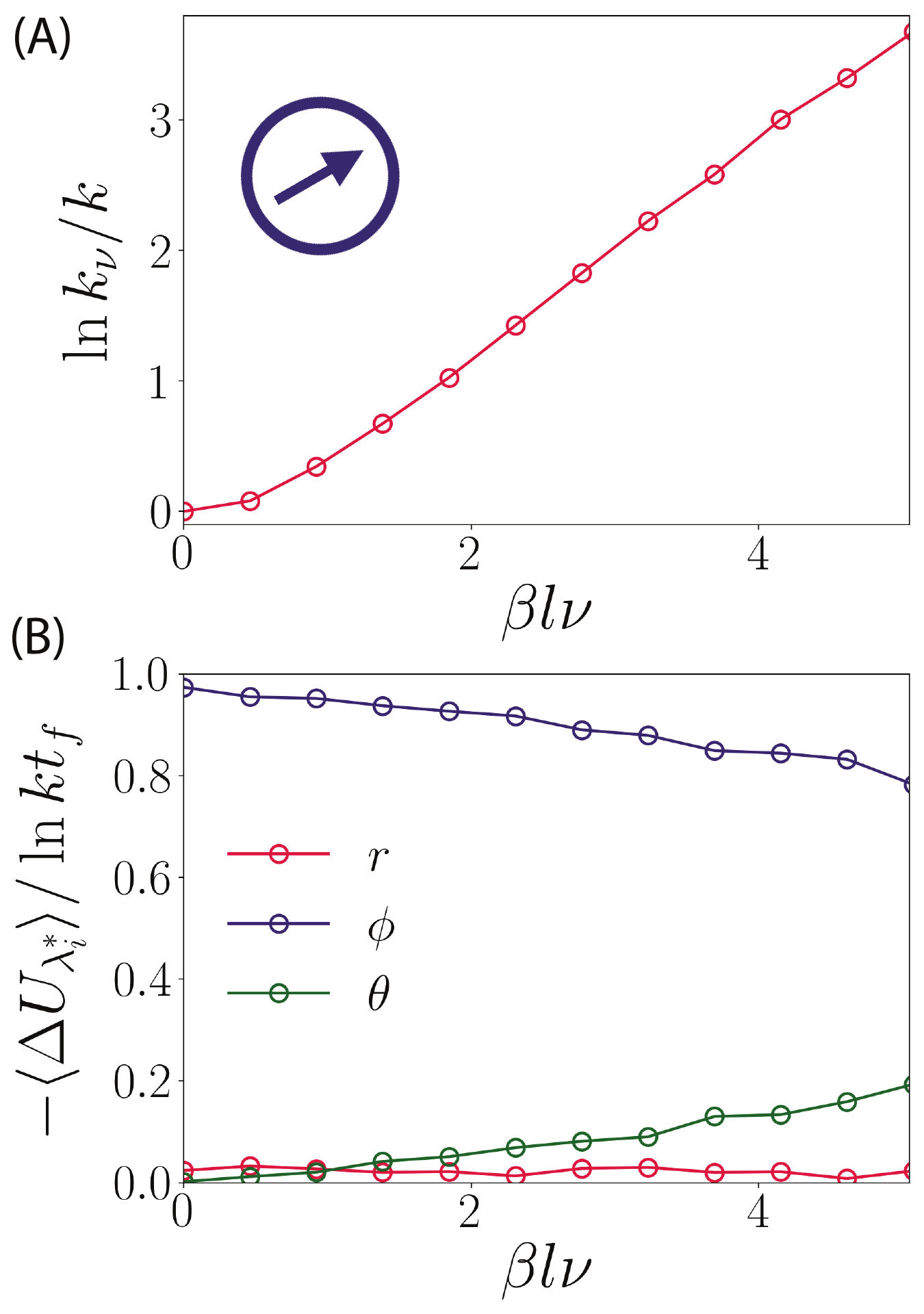}
    \caption{Kinetics of an active Brownian particle in an external potential. (A) The rate enhancement of the transition between the left and right wells as a function of activity. (B) Decomposition of the rate as a function of activity, in terms of polar coordinates $r=\sqrt{x^2+y^2}$, $\tan (\theta)=|y|/x$ and active director $\theta$.}
    \label{fig:2}
\end{figure}

We consider the rate to transition across the barrier over a range of driving amplitude between $\nu l/\kbt \in [0,5]$. We compute the rates by performing first mean passage time calculations from $6000$ trajectories initiated from the steady-state probability density within $A$ for each $\nu$. Figure \ref{fig:2} (A) illustrates the log ratio of the rates with respect to the equilibrium rate, $\nu=0$, as a function of the driving magnitude. The plot shows that the addition of activity enhances the rate of transition between the two states, and the form of the enhancement is non-linear. For small driving, the enhancement is observed to be quadratic with $\nu$, and with larger driving, the curve shows a turnover towards a linear regime.  

Using the decomposition in Eq.~\ref{eq:decomp}, and solving for the optimal driving force variationally, we find that over the range of $\nu$ considered, that despite the enhancement of the rate, the basic mechanism of the barrier crossing has not changed appreciably. This is observed in Fig. \ref{fig:2} (B). The importance of radial coordinate $r$ is found to be unchanged due to the addition of activity, and the importance of the director $\theta$ is observed to increase linearly with the Peclet, although even at the highest Peclet considered where the rate enhancement is approximately 2 orders of magnitude larger than the equilibrium rate, the total contribution is observed to be around 20\%. Across the whole range of driving magnitude considered, the dominant contribution to the rate is still the activation of the equilibrium reaction coordinate, $\phi$.

To understand the origin of the rate enhancement in Fig. \ref{fig:2} (A), despite the fact that the applied force is random, we can use the empirical observation that the mechanism does not change to deduce a simple model. In particular, we can use the separation of timescales to coarse-grain the continuous system into a  minimal discrete space Markov model. The model we consider has two spatial states, $A$ and $B$, reflective of the two metastable minima in the potential. In addition we imagine that within each spatial state, the particle has two internal states, denoted with $\theta$, one in which the active force is aligned with the equilibrium reaction coordinate for the $A$ to $B$ transition $\theta=0$, and one in which it is anti-aligned with the equilibrium reaction coordinate $\theta=\pi$. As a consequence of $\mathbf{\hat q_B} (\mathbf{r})=-\mathbf{\hat q_A} (\mathbf{r})$ in the full system, when the internal state is aligned with the equilibrium reaction coordinate for the $A$ to $B$ transition, it is anti-aligned with the $B$ to $A$ transition. Assuming that the diffusion of the active force over the course of a transition is negligible, the transition rates between the $A$ and $B$ states can be derived using Kramers-Langer theory~\cite{langer1969statistical,berezhkovskii2005one,peters2017reaction},
\begin{equation}\label{eq:lb1}
    k_{AB,\theta}(\nu)   =k e^{ \beta l \nu \cos(\theta)}
\end{equation}
while for the reverse transition 
\begin{equation}\label{eq:lb2}
    k_{BA,\theta}(\nu) = k e^{-\beta l \nu \cos(\theta)}
    \end{equation}
where $k=k_{AB,\theta}(0)=k_{BA,\theta}(0)$ is the rate to transition between the states in the absence of activity, and $\beta=1/\kbt$. Additionally, we model the transition between different active force alignments with the effective rate, $k_\theta=D_r \pi^2$, for $\theta=0$ to $\pi$ and $\theta=\pi$ to 0. Additional details for the derivation of these rates are shown in Appendix \ref{appendB}.

With these effective rates, we can construct and analyze the four state Markov model to coarse grain an active Brownian particle in a metastable potential.  The steady-state probability $p_i$ of each site $i=\{A(0), A(\pi), B(0), B(\pi)\}$ is given by
\begin{align}\label{eq:msm1}
    p_{A(\theta)} = \frac{k_\theta + k e^{-\beta l \nu \cos(\theta)}}{4k_\theta + 4k\cosh(\beta l \nu)}\notag\\
    %p_{A(\pi)}  =\frac{k_\theta + k e^{\beta l \nu}}{4k_\theta + 4k\cosh(\beta l \nu)}\notag\\
    p_{B(\theta)} = \frac{k_\theta + k e^{\beta l \nu \cos(\theta)}}{4k_\theta + 4k\cosh(\beta l \nu)}
    %p_{B(\pi)}  = \frac{k_\theta + k e^{-\beta l \nu}}{4k_\theta + 4k\cosh(\beta l \nu)}
\end{align}
and the corresponding total rate between states $A$ and $B$ summed over the two angles, can be computed analytically using the formalism of transition path theory for Markov jump processes~\cite{metzner2009transition}.  By computing the net flux and the probabilities, the rate of the driven system is related to the rate in equilibrium as,
\begin{align}
k_{\nu}  &=k \frac{k +  k_\theta  \cosh(\beta l \nu)}{ k_\theta + k \cosh(\beta l \nu)} \approx k \cosh(\beta l \nu)
\end{align}
where the second line follows under the assumption that $k \ll k_\theta$, or that the equilibrium rate is small compared to the rate for reorientation. 

\begin{figure}[t]
  \centering
    \includegraphics[width=0.49\textwidth]{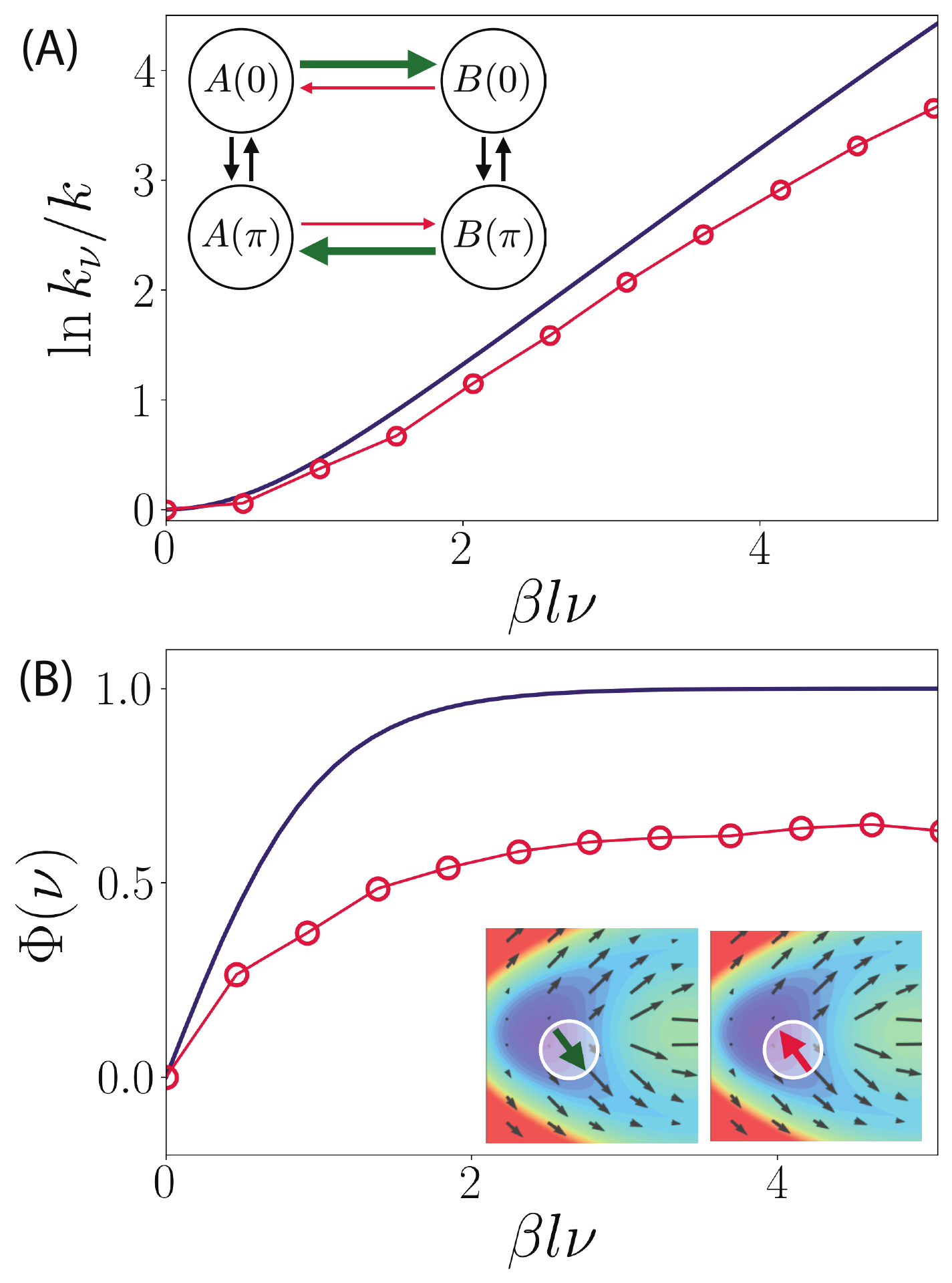}
    \caption{Mechanism of active rate enhancement. (A) Rate enhancement for the four state Markov model (shown in inset) as a function of activity (purple), along with the rate enhancement for the active Brownian particle in the external potential (red). (B) Alignment of the activity with the reactive flux for the Markov state model (purple) and active Brownian particle (red) in the external potential. Representative snapshots of states $A(0)$ and $A(\pi)$ are shown within the inset.}
    \label{fig:3}
\end{figure}

The effective rate from $A$ to $B$, $k_{\nu}$, for sample parameters $\ln k=-7$, $k_\theta \tau = 1$ and $\beta \nu l=1$ for time unit $\tau=1$ is shown in Fig. \ref{fig:3} (A) as a function of the dimensionless driving force $\beta \nu l$.   The same scaling behavior is observed as in the full two-dimensional model. A quadratic dependence on $\nu$ is found for small $\nu$ and a log-linear dependence is observed for larger $\nu$. The nature of the turnover of the rate enhancement under the barrier crossing regime from quadratic to linear is closely linked to the competition between the pathways that the system takes during the transition. This can be measured by evaluating the alignment of the active force with reactive flux. 
For the $A$ to $B$ transition, there are two available pathways within the Markov model $A(0)$ to $B(0)$ where the active force is aligned with the reaction coordinate, and $A(\pi)$ to $B(\pi)$ where it is anti-aligned. Defining the alignment of the active force with reactive flux as $\Phi$, for the Markov model
\begin{align}
\Phi &= \frac{\sum_{\theta=0,\pi} p_{A(\theta)} k_{AB,\theta}\cos(\theta) }{\sum_{\theta=0,\pi} p_{A(\theta)} k_{AB,\theta} } \notag \\
    &= \frac{k_\theta \sinh(\beta l \nu)}{k + k_\theta \cosh(\beta l \nu)} \approx \tanh (\beta l \nu)
\end{align}
where $p_{A(0)} k_{AB,0} $ and $p_{A(\pi)} k_{AB,\pi}$ are the fluxes through pathways $A(0)$ to $B(0)$, and $A(\pi)$ to $B(\pi)$ respectively. For the final form, we took the limit of $k/k_\theta \rightarrow 0$, which is valid for the high barrier limit.  This function is plotted in Fig.~\ref{fig:3}(B).

From $\Phi$, the mechanism of rate enhancement can be understood. Around equilibrium, the reaction takes place through all possible pathways, some of which that are enhanced and some that are suppressed due to addition of active forces. For the active Brownian motion, the alignment of the director with the reaction coordinate averages to 0. As the activity increases, the system reacts more selectively through the enhanced channel because the flux through that pathway is increased. At some finite driving strength, the system only reacts through the enhanced pathway, as its flux is exponentially increased while the other is exponentially suppressed. The region where only one dominate pathway is accessed  corresponds to the regime where the enhancement scales linearly with the onset of activity. Hence, the quadratic to linear trend with increasing activity is a signature of broken detailed balance, and highlights a simple mechanism through which active forces perturb reactions. 

We can evaluate the same order parameter that measures the alignment of the active force with the reaction coordinate in the full two-dimensional system given knowledge of the equilibrium reaction field, $\mathbf{\hat q_B}$. For the continuous system this is given by
\begin{equation}\label{eq:op}
\Phi(\nu)=\langle \mathbf{\hat e} \cdot \mathbf{\hat q_B} \rangle_{B|A, \nu}
\end{equation}
where $\mathbf{\hat e}=\{\cos(\theta),\sin(\theta)\}$ is the direction of the active force, $\mathbf{\hat q_B}$ is evaluated for $\nu=0$ and the average is taken within the reactive path ensemble at different values of $\nu$. The same qualitative behavior is observed as in the discrete system, and illustrated in Fig.~\ref{fig:3}(B) along side the discrete model result. Near equilibrium the fraction of the reactive flux enhanced by activity increases linearly with $\nu$, before plateauing for larger values of $\nu$. As in the discrete system this plateau corresponds to the transition between quadratic to linear rate enhancement. Certainly, for very large values of $\nu$ distinct reactive channels will generically open, and the enhancement of the rate with activity will have less to do with preferentially increasing the flux along the equilibrium reaction field.  Nevertheless, for this system we have clarified an example of how random forces can sped up barrier crossing events by aligning with a reference equilibrium reaction field.

\subsection{Rate enhancement in an interacting system}
\begin{figure}[t]
  \centering
    \includegraphics[width=0.49\textwidth]{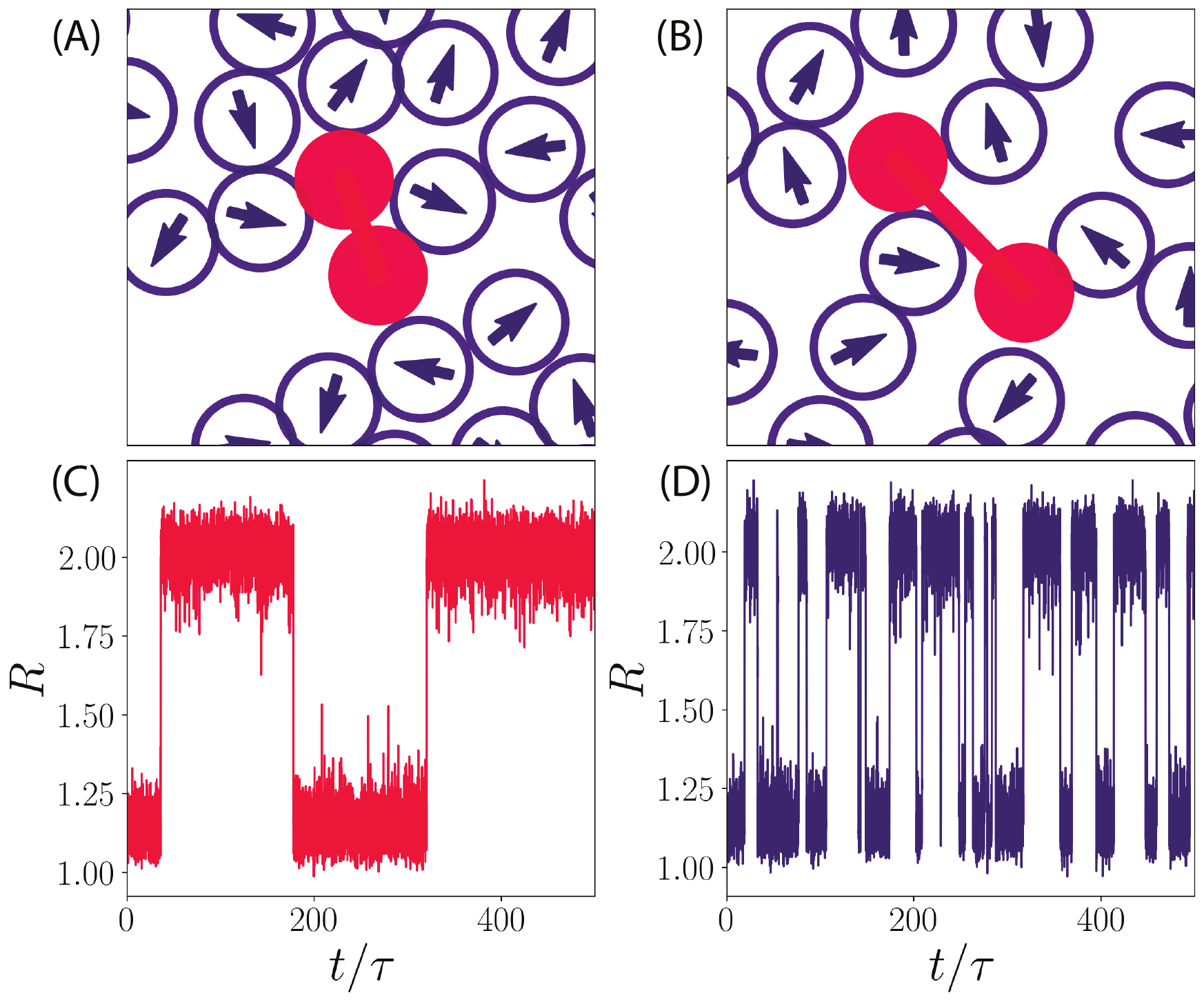}
    \caption{Kinetics of a passive dimer in an active bath. Representative snapshots of the system in its (A) reactant state and the (B) product state. Timeseries of the order parameter $R$ corresponding to the distance between the two monomers (C)  at equilibrium, $\nu=0$ and (D) at finite activity $\beta \nu \sigma =16$.}
    \label{fig:4}
\end{figure}

The analysis and arguments in the previous section are general, and suggest that rate enhancement through the application of active forces should be present in many systems. To test this idea, we consider the dimerization of a passive dimer in an active bath where the two states are visualized in Figs. \ref{fig:4} (A) and (B). This system has been been previously studied to demonstrate how variational path sampling can be used to evaluate accurate estimates of out-of-equilibrium reactive processes~\cite{das2022direct}, and is motivated to understand how an active bath can enhance passive reactive processes such as self-assembly, and convert chemical work to mechanical work~\cite{mallory2020universal,sokolov2010swimming}. The system comprises a passive dimer and 78 active Brownian particles interacting via volume exclusion that is parameterized through a Weeks-Chandler-Anderson potential~\cite{weeks1971role}
\begin{equation}
    V^{\mathrm{WCA}}(\mathbf{r}) = 4\epsilon \left( \left( \frac{\sigma}{r} \right)^{12}   - \left( \frac{\sigma}{r} ^{12} \right) + \frac{1}{4} \right) \Theta(r_{\mathrm{WCA}} -r)
\end{equation}
with energy scale $\epsilon$ and particle diameter $\sigma$, truncated at $r_{\mathrm{WCA}} = 2^{\frac{1}{6}} \sigma$. The solvent particles feel an added drift from the director given by the form
\begin{equation}
    \mathbf{F}^a  = \nu \mathbf{\hat e}
\end{equation}
where for each particle, $\mathbf{e}_i = (\cos(\theta_i),\sin(\theta_i))$ and $\nu$ is the strength of the active self-propulsion. The dimer feels an additional double well potential of the form
\begin{equation}
    V^{\mathrm{DW}} = \Delta V [1 - (R - r_{\mathrm{WCA}} - 0.45\sigma)^2/(0.45\sigma)^2]^2
\end{equation}
where $R$ is the distance between the monomers and $\Delta V$ is related to the height of the barrier. This form of the potential leads to two metastable states centered at $r_{\mathrm{WCA}}$ and $r_{\mathrm{WCA}} + 0.9 \sigma$, and we use the following indicator functions to identify the two states
\begin{equation}
    h_A(R) = \Theta(-R+1.25\sigma) \qquad  h_B(R) = \Theta(R-1.85\sigma)
\end{equation}
and study the reaction from collapsed state denoted by $A$ to the extended state denoted by $B$. We chose the density of the system to be 0.6 $\sigma^{-2}$, $\epsilon = \gamma = \sigma = 1$, $\kbt = 0.5$, $\Delta V/\kbt = 14$, reduced time $\tau = \sigma^2 \gamma / 2\kbt = 1$ and trajectory path time $t_f = 0.2$. The friction coefficient for the dimer and solvent particles is set $\gamma_d = 2\gamma$ and $\gamma_b = \gamma$ respectively, and the rotational diffusion constant for the director is set to $D_\theta = 1/\tau$. We use a timestep $dt = 10^{-4}\tau$ for integrating the equations of motion.

In a previous study, it was found that the  transition from the collapsed to the extended state was observed to be enhanced by multiple order of magnitudes under the addition of activity to the solvent particles.~\cite{das2022direct} This can be inferred from Fig. \ref{fig:4} (C) and (D) that shows the timeseries of the bond distance $R$ at equilibrium  $\nu=0$ and a finite $\beta \nu \sigma=16$, where the transitions in the later is significantly more frequent.  In equilibrium, the dominant degrees of freedom that mediate the reaction of the dimer are just its bond distance. We are interested in understanding how the reaction mechanism changes under the addition of activity that is applied non-locally through the solvent. To do so, we use variational path sampling~\cite{das2022direct}, to deduce the importance of individual degrees of freedom through the optimization of a many-body time-dependent committor~\cite{singh2023variational}. The method involves a choice of a set of descriptors as well as an ansatz to represent the time dependent committor, and the optimization of a variational loss given by the change of action defined in Eq. \ref{eq:OM} and averaged within a reactive path ensemble. Post optimization, the loss function can be decomposed into contributions from different descriptors, following Eq.~\ref{eq:decomp}, offering a way to gain higher order insights into the reaction. 

For this system, we consider the reaction under a range of nonequilibrium activity given by $\beta \nu \sigma = [0,2,4,6,8,10,12,14,16,18,20]$. For each $\nu$, we obtain approximately 200-300 reactive trajectories from brute-force simulations propagated using the Euler-Muruyama discretization scheme, and store the trajectories along with the noises. To describe the system we utilize a set of  parameters that describe intra-dimer interactions as well as the solvent-dimer interactions. We consider the bond distance between the dimer $R$ as one of the order parameters. To encode solvent effects, we rotate the system along the axis parallel to the bond distance and consider the distance of the solvent from the center of mass of the dimer $R_s$, and the angle between the solvent and the bond vector, denoted $\phi_s$. Finally, the effect from the solvent directors is encoded using the relative angle $\theta_s$ between the solvent director and the distance vector between the solvent and one of the monomers. This is done for a total of twelve solvent particles that are closest to the center of mass of the dimer at a given configuration, which we later find is sufficient to describe the mechanism of the reactive event. Details of the optimization and the form of the ansatz is discussed in Appendix \ref{appendC}.
% and the code for optimization as well as validation is available on Github~\cite{github}.

\begin{figure}[t]
  \centering
    \includegraphics[width=0.49\textwidth]{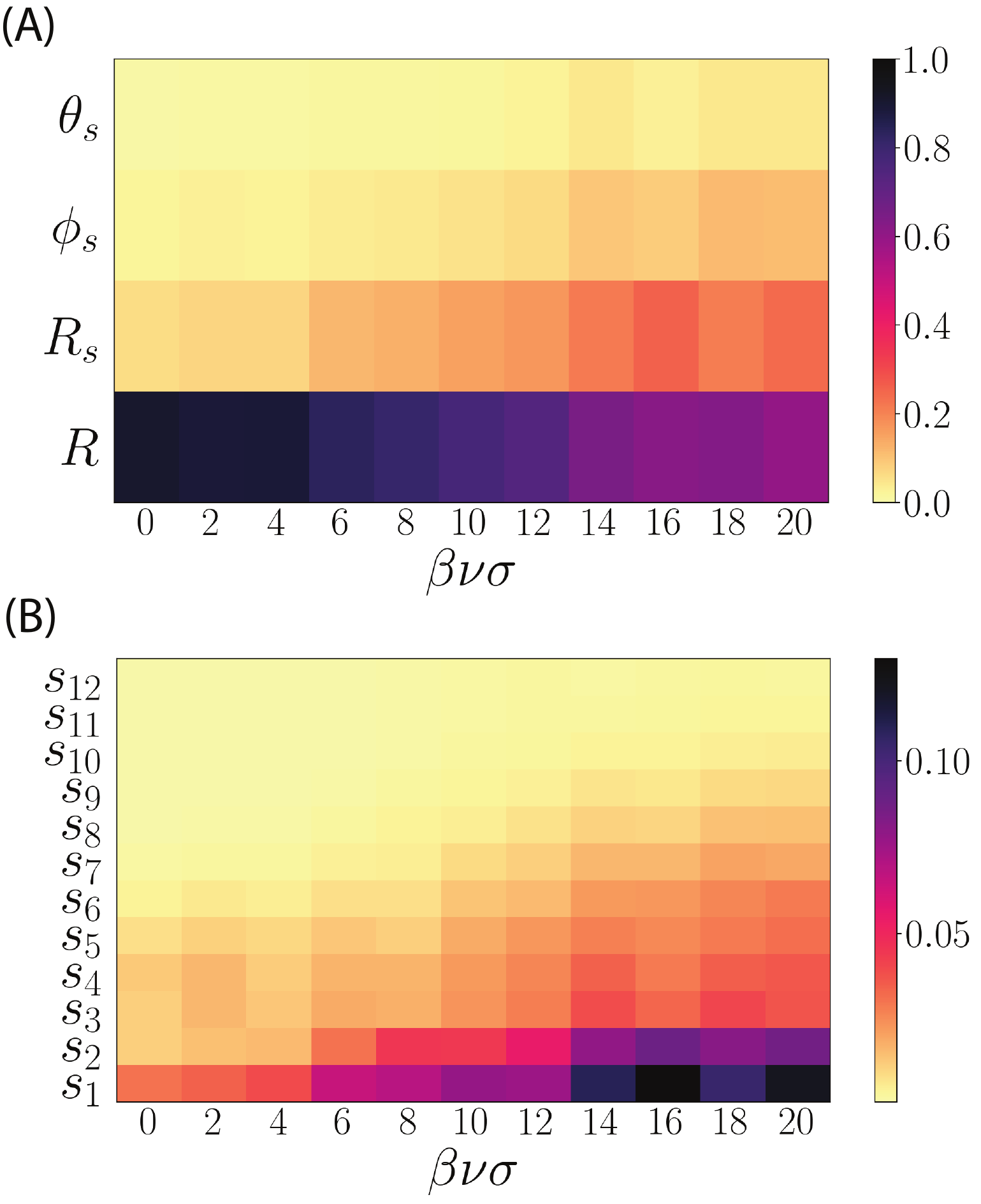}
    \caption{Change in the reaction mechanism due to addition of activity to the solvent particles. (A) Importance of different descriptors to the reactive event as a function of the activity. The contributions of the solvent descriptors relative to  $\ln k(\nu) t_f$, computed by summing over the individual contributions of the 12 solvent particles closest to the center of mass of the dimer. (B) Individual contributions of the solvent particles relative to  $\ln k(\nu) t_f$ computed by summing over the contributions from the solvent descriptors as a function of the $\nu$. The solvent labels are assigned according to the distance from the center of mass of the dimer. }
    \label{fig:5}
\end{figure}

Post training, we consider the decomposition of the change in action into effective contributions from the different order parameters, quantified with $\Delta \tilde U_{\lambda^*} ^{j}$. 
These components are first computed for the order parameters $R$, $R_s$, $\phi_s$ and $\theta_s$, where the contribution from solvent descriptors are summed over all the 12 nearest solvent particles. The importance of these descriptors as a function of the $\nu$ is plotted in Fig. \ref{fig:5} (A). The figure shows a monotonic decrease in the effective importance of the dimer distance $R$ as the nonequilibrium driving is increased -- in equilibrium, the dimer distance is an excellent descriptor of the reaction coordinate, with its component contributing to about 90 \% of the total change in action, and this eventually decreases to approximately 60 \% for the highest $\nu$. Consequentially, the solvent descriptors become more important, and an increase in the importance of the solvent distances $R_s$ is observed for medium driving. For large $\nu$, the relative angle of the solvent contributes significantly, followed by an emergence in the importance of the relative angle of the director. 

We also consider the change in importance of the solvent particles as a function of the distance from the center of mass of the dimer. To do so, we sum up the effective contributions of the change in action from $R_s$, $\phi_s$ and $\theta_s$ of each solvent particle individually and plot it as a function of nonequilibrium activity in Fig. \ref{fig:5} (B). The solvents are indexed by their distance from the center of mass of the dimer, with 1 corresponding to the closest solvent particle to the dimer. For all $\nu$, there is a monotonic decrease in the importance of the solvent particles as the distance of the solvent from the center of mass of the dimer increases. This indicates that the form of the architecture for the neural network was successful in capturing solvent effects that is expected to decay with increasing distances. In terms of the trends with respect to $\nu$, we find the closest solvent particles to contribute minimally to the reaction at equilibrium, and their relevance is observed to become significant for high activity. What is more striking is the long-range effect that is induced by nonequilibrium driving. We find distant solvent particles to become important for high driving, which we find is due to the interference with the backward reaction that involves solvent particles behind the monomers pushing monomers together. 

\begin{figure}
  \centering
    \includegraphics[width=8.5cm]{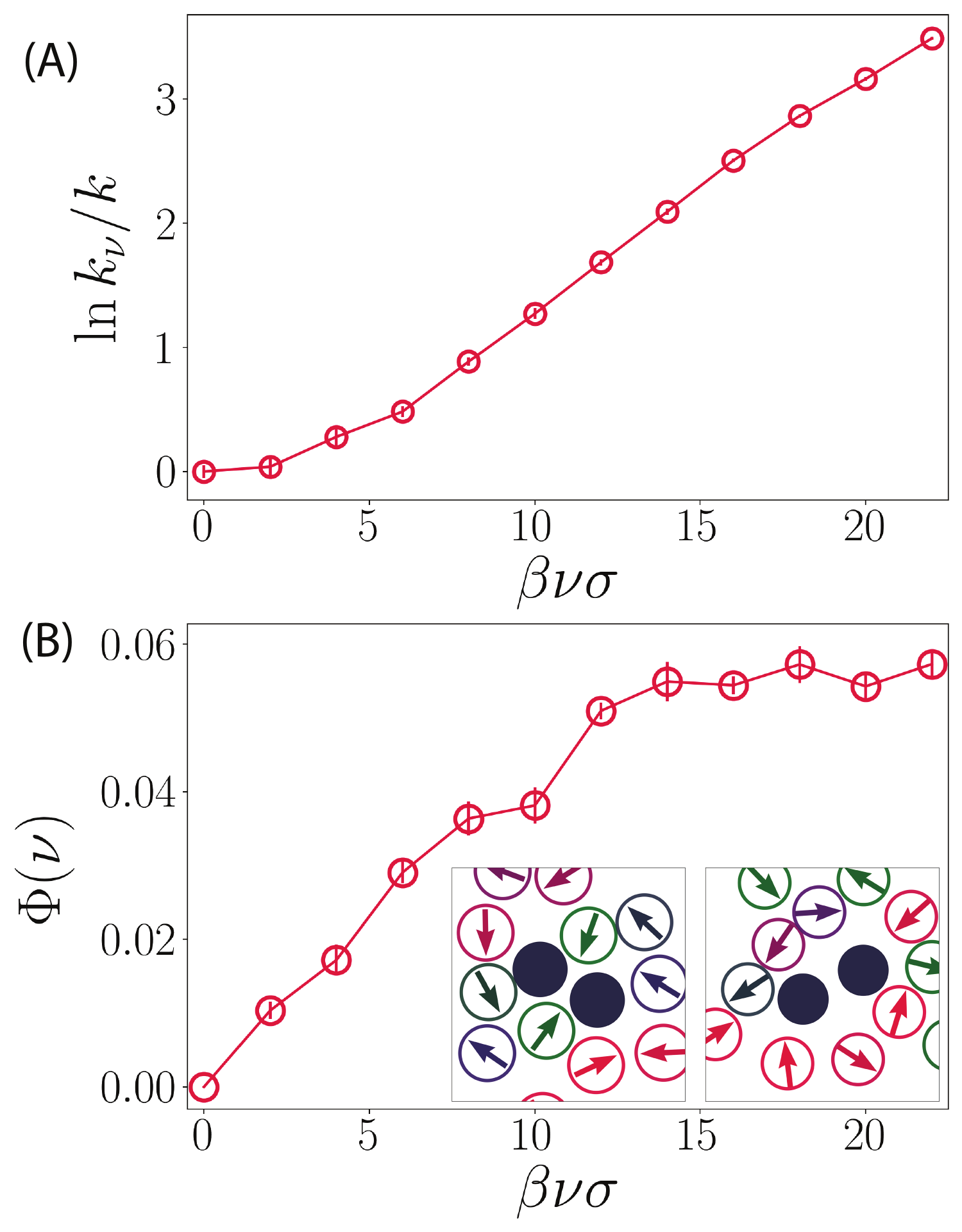}
    \caption{Mechanism of rate enhancement for the passive dimer in active solvent. (A) Log rate enhancement of as a function of the activity (B) Flux weighted alignment of the active forces with the reaction field computed using Eq. \ref{eq:phi_dimer}. The inset shows representative snapshots for the sub states that have the directors aligned towards the reaction field (left) and against the reaction field (right). The solvent particles are colored according to the alignment of their director with the direction of the optimal control force for the equilibrium reaction, with red, purple and green denoting antiparallel, perpendicular and parallel alignment. }
    \label{fig:6}
\end{figure}

The considerably large increase in importance of the solvent molecules under the onset of nonequilibrium driving for this system indicates that the reaction is strongly mediated by fluctuations of the polarization of the solvent. Remarkably, the trends observed for the rate enhancement between the two-dimensional system and the passive dimer are the same - quadratic at smaller driving, linear at larger driving. This is observed in Fig.~\ref{fig:6}(A). Additionally, as in the active particle in an external potential, we can compute the alignment of the active force with the reaction field, which for the many particle case is
\begin{equation}\label{eq:phi_dimer}
\Phi(\nu) = \frac{1}{\sqrt{N_s}}\left \langle \sum_{i=1}^{N_s} \hat{\mathbf{e}}_i \cdot \frac{\nabla_i q_B}{|\nabla q_B| } \right \rangle_{B|A,\nu}
\end{equation}
where $N_s$ are the number of solvents in our description of the committor.  Here we consider the reaction field deduced from the variationally optimized committor in equilibrium, which while largely determined by $R$ has some contribution from the solvent. This provides an oppertunity for the active force to align with that component of the reaction field.  
 The alignment of the activity with the equilibrium reaction field for the dimer system is shown in Fig.~\ref{fig:6}(B), illustrating that just as in the simpler cases, the rate enhancement is due to the preferential amplification of specific transition pathways, those for which the net activity aligns with the reaction coordinate. In this case, the alignment of the active force with the reaction coordinate is essentially the alignment of the polarization field of the active Brownian particles in the vicinity of the dimer, with a vector perpendicular to the dimer bond vector. The mechanical interpretation being that the enhanced flux of the conformational change occurs because the active particles pry the dimer apart. Near equilibrium, the active force also pushes the dimer back together, suppressing the rate, but the flux associated with this pathway is exponentially suppressed at large active force, $\nu$.

\subsection{Why do active forces enhance kinetics?}

We conclude by concretely answering why active forces tend to accelerate kinetics. The effect of external forces on the kinetics of multidimensional potential energy surfaces can be understood by considering Doob's picture of conditioning, as embedded in the reweighting relations at its variational form in Eqs. \ref{eq:OM} and \ref{eq:rate_bound}~\cite{singh2024splitting}. This states that there exists a force related to the committor that encapsulates all reactive fluctuations. Specifically, the force can be decomposed into a unit vector field and a scalar field, and in the limit that the initial equilibrium barrier persists, the affect of external forces can be understood by the alignment of these forces with the reaction field. Within this context, active forces are unique in that their application to passive systems enhance some transitions and suppress others. Due to the exponential coupling of rates with external forces, the addition of equally suppressed and enhanced pathways leads to an effective enhancement of rate and collapse of the path ensemble onto the pathway that is sped up. The form of the rate enhancement has a unique trend that reflects this change in ensemble. At small driving, the rate enhancement scales log-quadratically due to the competition between the suppressed and the enhanced pathways. At large driving, the reaction takes place selectively through the enhanced pathways leading to a log-linear enhancement. While we believe that some of the results can be inferred from equilibrium based methods involving rescaling of the temperature, or using instantonic methods under the small noise limit, the complete mechanism of these phenomena, especially when considering generality to complex interacting systems requires consideration of more than one pathway, and of how detailed balance breaks down under the addition of activity.

\section{Emergent metastability}
%For systems evolving within nonequilibrium steady-states, their probability distributions depend on both energetic as well as kinetic details. While at equilibrium, detailed balance ensures that there is no net probability flux between any two states and as a consequence the steady-state distribution depends only on relative energetics. 
Nonequilibrium steady-states generically have persistent currents through configurational space. Steady-state probability densities depend on both local energetics, but also the specific rates at which probability enters or leaves a particular configuration. 
A range of studies have shown that nonequilibrium driving can lead to emergent stability of states that are thermodynamically unfavorable in equilibrium. Motility induced phase separation in repulsive active matter is one canonical example, where dynamical trapping stabilizes a condensed state~\cite{cates2015motility}. Additionally, the phenomena of emergent stability due to nonreciprocal interactions has been widely studied within active matter, self assembly and biophysics~\cite{saha2020scalar}. Often the emergent state can be metastable, and mechanistic analysis into the transitions between these states can be challenging due to the breakdown of detailed balance, and the emergence of persistent probability currents that can mediate the reaction. 

%% Flux balances and breakdown in time reversal symmetry
As we have shown in the previous section, the addition of a non-conservative force can enhance the reactive flux through some pathways, and suppress others. This asymmetry can be harnessed to establish metastability through kinetic trapping. For transitions between two states, stability can be enhanced provided the rate into one of the states is enhanced while the transition out of the state is suppressed. However, if the transitions occur through only one available channel then the effective bistable system is equilibrium-like in that the mechanisms as encoded in the commitor and optimal force will be simply related to each other. Specifically, if $q_B(\mathbf{r})$ is the commitor for the transition $A$ to $B$, and $q_A(\mathbf{r})$ the commitor for the reverse transition, then $q_A(\mathbf{r})=1-q_B(\mathbf{r})$. As a consequence, the stability can be mapped to an effect energetic stabilization. However, if there are multiple channels available to transition, and one is enhanced in one transition direction while the other is enhanced in a different one, then metastability can emerge from the pinning of probability currents. This mechanism for kinetic stability is not possible in equilibrium. Using our ability to decode reactive processes within nonequilibrium steady-states, in this section we consider the stabilization of thermodynamically unstable states by the breaking of microscopic reversibility.  

%General approaches for understanding emergent behavior
\begin{figure}[t]
  \centering
    \includegraphics[width=0.48\textwidth]{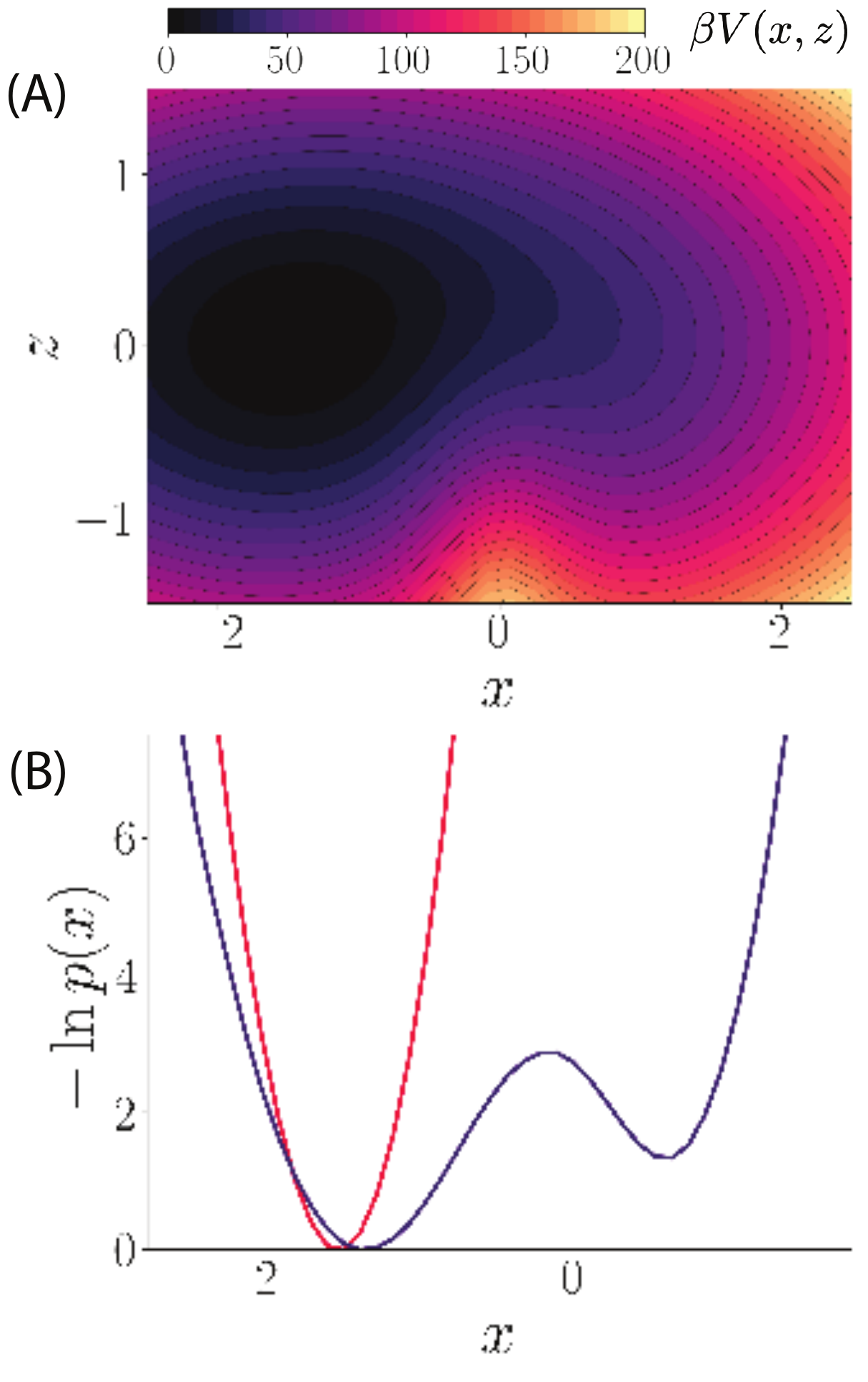}
    \caption{(A) Two-dimension potential, $V(x,z)$ defined in Eq.~\ref{eq:shearpot}. (B) Marginal probability in equilibrium $\dot{\gamma}=0$ (blue) and under shear $\dot{\gamma}=90$ (red).}
    \label{fig:7}
\end{figure}

%Insights from the Crooks-Kawasaki relationship/ Owen, Horowitz & Gingrich
\subsection{Detailed balance violations and metastability}
To explore the interplay between the breakdown of detailed balance and emergent metastability, we first consider a minimal two dimensional system with a harmonic potential along variables $x$ and $z$, and a small nonlinearity that adds an additional repulsive force for motion along $x$. The total potential energy is given by
\begin{eqnarray}\label{eq:shearpot}
     V(x,z)/\kbt &=& 40 (x-1)^2 + 10z^2 +\\
    &&\frac{1}{0.03x^2 + 0.01(z+1.5)^2 + 0.01} \notag
\end{eqnarray}
and a contour plot of the potential energy surface is shown in Fig. \ref{fig:7} (A). The stiffness of the two springs in $x$ and $z$ result in a single potential minima located near $z=0$ and $x=-1.5$. The marginal distribution, $p(x)=\langle \delta [x-x(t) ]\rangle$, shown in Fig. \ref{fig:7} (B), is computable from integrating over the resultant Boltzmann distribution. There is little signature of the non-linearity, and $p(x)$ is a strictly convex function.

To drive the system out of equilibrium, we consider the addition of a shear force to the particle moving in $ V(x,z)$. In particular, we study the overdamped dynamics with non-conservative drift in the $x$ direction $F_x = -\partial_x V(x,z) + \dot \gamma z$, that takes the form of a linear shear profile with rate $\dot{\gamma}$. In the $z$ direction $F_z = -\partial_z V(x,z)$. Using $\gamma=\kbt=1$, and evolving the dynamics using an  Euler-Murayama integrator with discretization $dt = 0.004\tau$, we find that at a critical value of the shear rate, $\dot{\gamma}=80$, the marginal distribution, $-\ln p(x)$, becomes non-convex. In particular, we find that the shear stabilizes a metastable state with local minima at $x=0.5$. For a range of shear rates above this threshold, the system is bistable. The underlying potential does not support this metastability, thus it is directly the result of the non-conservative force.

We find that for a range shear rates for which $\ln p(x)$ is bimodal, the system transitions between two locally stable states infrequently, affording a separation of timescales between local relaxation and transition, allowing us to define rates and analyze the mechanism of the transitions. 
Defining indicator functions
\begin{equation}
    h_A(\mathbf{r}) = \Theta(-x-0.5), \qquad h_B(\mathbf{r}) = \Theta(x-0.2)
\end{equation}
we can compute rates and deduce the optimal control forces as a function of shear rate. We find that coincident with the formation of a new stable state is a decoupling of the transitions between trajectories that move the system from $A$ to $B$ or back. As shear increases the mechanism by which a transition occurs from $A$ to $B$ is characterized by motion along the diagonal in the $x-z$ plane with equal contributions to the activation in both. However, the mechanism by which the system transitions from $B$ to $A$ is increasingly different with increasing shear, characterized by larger contributions along the $x$ direction relative to the $z$ direction. The distinction between the pathways the system transitions is a specific consequence of the shear forces that act to convect the particle, and a general manifestation of the breakdown in microscopic reversibility. 

\begin{figure}[t]
  \centering
    \includegraphics[width=0.48\textwidth]{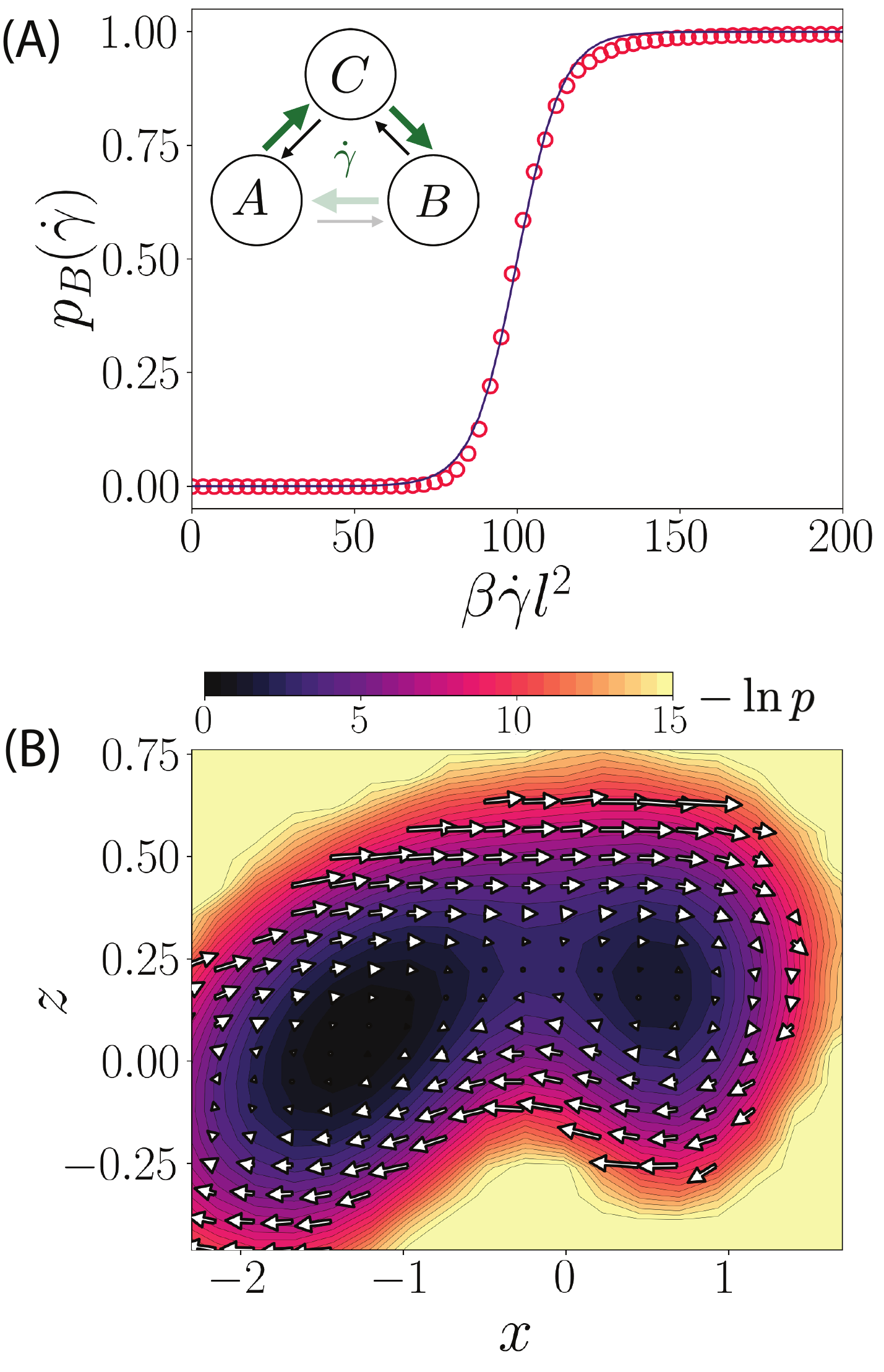}
    \caption{A mechanism of stability due to nonconservative forces (A) Enhancement of the probability of state $B$ as a function of the reduce shear rate $\beta \dot \gamma$ (red), along with the probability enhancement predicted from Markov state model in Eq. \ref{eq:pb} with fitting parameters $l=0.27$ and $\epsilon$=14.54. (purple) (B) Log probability density and the velocity field defined in Eq. \ref{eq:vf1} for the 2D system at shear $\beta \dot \gamma l^2=90$.}
    \label{fig:8}
\end{figure}

To understand the origin of the metastability, we can coarse-grain the system into a minimal discrete space Markov model. The model we consider has three spatial states, $A,B$ and $C$. We imagine that $B$ and $C$ are energetically unfavorable relative to state $A$, by an energy $\beta \epsilon>1$, such that in equilibrium, the probability is concentrated nearly entirely in state $A$. This reflects the single stable state in the absence of the shear in the two-dimensional model. In equilibrium, transitions between $A$, $B$ and $C$ obey detailed balance. To model the nonlinearity in the potential that impedes motion along the $x$ direction at small values of $z$, we assume that transitions between $A$ and $C$ and also $C$ and $B$ have a prefactor $k_0$, while transitions between $A$ and $B$ have a prefactor $k_0 \nu$ with $\nu \ll 1$. Finally to model the shear, we invoke analogous assumptions as used in the previous section to add a drift that favors cycles $A \rightarrow C\rightarrow B$, and disfavors cycles $A \rightarrow B\rightarrow C$ with affinity $\beta \dot{\gamma} \ell^2$ with $\ell$ a characteristic lengthscale. Under these assumptions the rates between states $A$ and $C$ are
\begin{equation}
k_{AC} = k_0 \frac{e^{-\beta \epsilon+\beta \dot{\gamma} \ell^2 /2}}{1+e^{-\beta \epsilon}} 
\qquad
k_{CA} = k_0 \frac{e^{-\beta \dot{\gamma} \ell^2 /2}}{1+e^{-\beta \epsilon}}
\end{equation}
those between $B$ and $C$ are
\begin{equation}
k_{CB} = k_0  e^{\beta \dot{\gamma} \ell^2 /2}
\qquad
k_{BC} = k_0  e^{-\beta \dot{\gamma} \ell^2 /2}
\end{equation}
and finally those between $A$ and $B$ are
\begin{equation}
k_{AB} = \frac{k_0 \nu e^{-\beta \epsilon}}{1+e^{-\beta \epsilon+\beta \dot{\gamma} \ell^2 /2}} 
\qquad
k_{BA} =  \frac{k_0 \nu e^{-\beta \dot{\gamma} \ell^2 /2}}{1+e^{-\beta \epsilon}} 
\end{equation}
with the connectivity and driving 
 illustrated in Fig.~\ref{fig:8}. 

This minimal Markov model allows for transitions between $A$ and $B$ to occur through two channels. There is a direct path that is slow at equilibrium, due to $\nu \ll 1$, and an indirect path through $C$. In equilibrium, in the absence of the shear, for $\nu < e^{-\beta \epsilon}$ the latter dominates. Because of microscopic reversibility, it is the dominate path for both $A$ to $B$ as well as $B$ to $A$. Out of equilibrium, for finite $\dot{\gamma}$, microscopic reversibility is broken, and there exists a net cycle through the state space. Reactive transitions are likely to follow the steady-state probability flux, and so the path that takes $A$ to $B$ continues to proceed through state $C$, while the reverse, transitions that take $B$ to $A$, increasing follow the direct route. Because the direct route is slow, the circulating flux can cause probability density to build up on state $B$. In the limit that $\nu \rightarrow 0$, the probability of state $B$ is 
\begin{equation}\label{eq:pb}
p_B(\dot{\gamma}) \approx \frac{1}{1+e^{-\beta \dot{\gamma}\ell^2}+e^{-\beta (2 \dot{\gamma}\ell^2-\epsilon)}}
\end{equation}
and in plotted in Fig.~\ref{fig:8}(A) and approaches 1 in the limit of large $\dot{\gamma}$. As $\nu \rightarrow 1$, there is still in increase in the probability of $B$ with $\dot{\gamma}$, but it is not large and $A$ remains the stable state. We can conclude that the mechanism of the increase in probability of $B$ is due to the existence of the bottle neck transition that becomes the dominate reaction channel out of equilibrium, that serves to pin probability flux. 

The minimal model suggests that the mechanism for the emergence of the metastable state derives from the alignment of transitions with the driving forces for probability flux in the steady-state. We can quantify this in the two-dimensional model by evaluating the
Fokker-Planck velocity field~\cite{boffi2024deep} defined as
\begin{equation}\label{eq:vf1}
\mathbf{v}(x,z) = \frac{1}{\gamma} \mathbf{F}(x,z) -\frac{\kbt}{\gamma} \nabla \ln p(x,z)
\end{equation}
which is shown in Fig.~\ref{fig:8} (B). With $\dot{\gamma}=90$, we find clear evidence of currents that circulate probability from the left well to the emergent right well, which are pinned, or deflected at the location of the repulsive force near $x=0$. Measuring the accumulated probability in the right well, as $p_B=\langle h_B\rangle$ as a function of shear, we find that its function form is well reproduced by the Markov model, shown in Fig.~\ref{fig:8}(A).

\subsection{Stabilization of an unfolded \\polymer under shear}
The analysis and arguments in the previous section are general, and suggest that the breaking of time-reversal symmetry by non-conservative forces can lead to the stabilization of otherwise unstable states.  To test this idea, we consider a specific example of a system that exhibits metastability far from equilibrium, a polymer grafted on a repulsive wall under the action of a shear force. These system has been studied extensively over the last few decades due its fundamental and practical relevance to biological systems~\cite{smith1999single,perkins1995stretching,perkins1997single,jaspe2006protein,huang2010semidilute,doyle2000dynamics}. A particularly relevant example includes simulations of grafted polymers under bad solvent conditions that have been employed to gain insight into the role of the protein von Willebrand factor in the process of blood clotting.~\cite{alexander2006shear,schneider2007shear,winkler2006semiflexible,siediecki1996shear} These polymers have been observed to exhibit a globule-stretch transition at a critical shear, similar to the unfolding of the von Willebrand factor into thin fibers under physiologically relevant shear rates.~\cite{buguin1996unwinding,alexander2006shear} The breakdown of detailed balance coupled with the long relaxation timescales of the polymer makes a mechanistic study of the unfolding-folding transition intractable using standard kinetic analysis. Consequentially, most of the work on this system has primarily focused on investigating the static properties with the introduction of shear. Here we first investigate the mechanistic differences between the folding and the unfolding transition to demonstrate how reactions between metastable states that emerge out of equilibrium can be distinct from those in equilibrium, and in doing so, we also highlight the mechanism through which probability currents can stabilize thermodynamically unfavorable states.

\begin{figure}[t]
  \centering
    \includegraphics[width=0.48\textwidth]{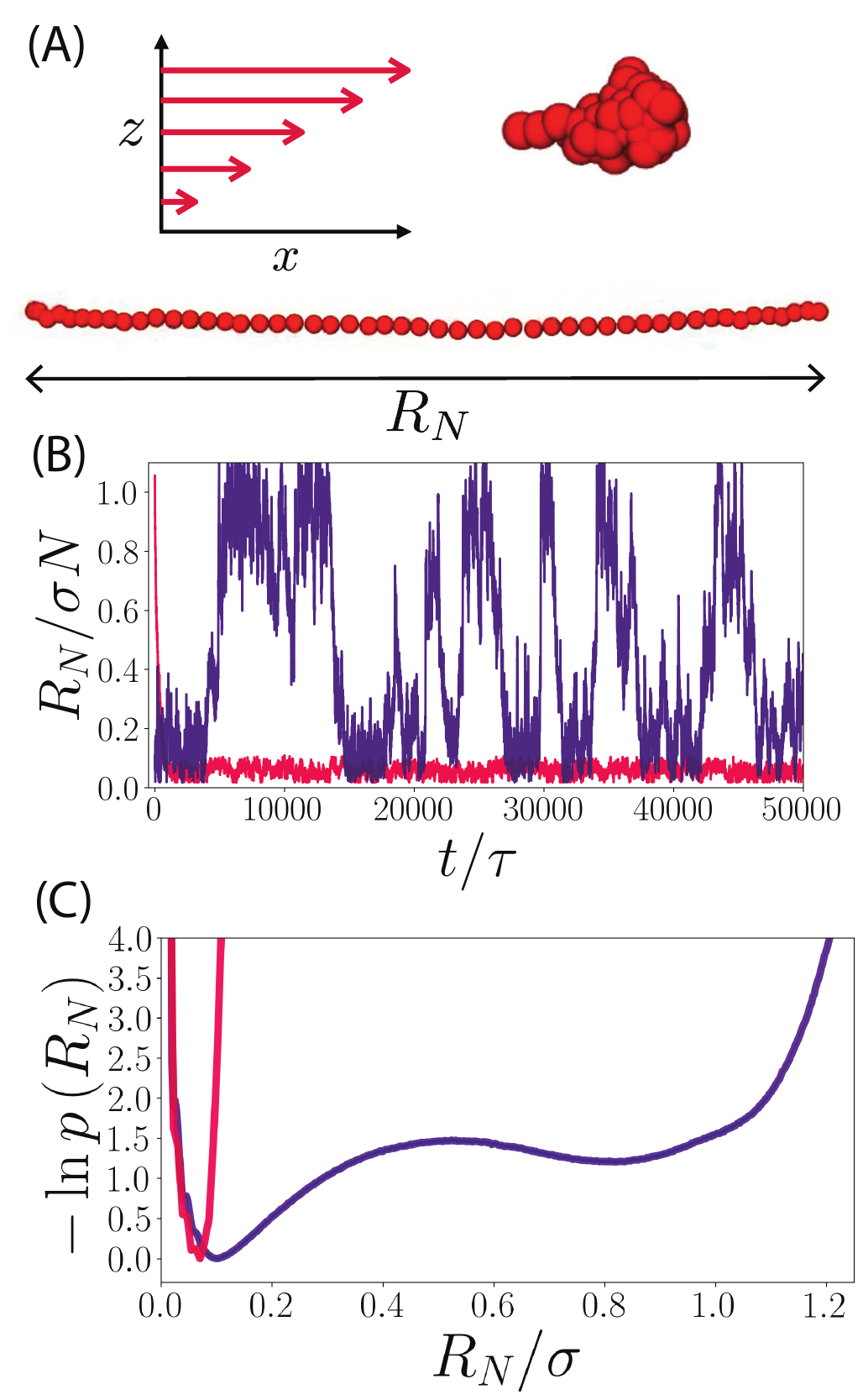}
    \caption{Shear induced metastability of a polymer grafted on a repulsive wall. (A) Representative snapshot of the polymer in the globular and stretched state. (B) A time series of the reduced end to end distance $R_N/N \sigma $ of the polymer at equilibrium (red), initiated from the stretched state showing immediate relaxation to the globular state. At finite shear (purple), the polymer is observed to transition between the two states exhibiting metastability. (C) The log density along the end to end distance for the polymer at equilibrium (red) and at finite shear (purple).}
    \label{fig:9}
\end{figure}

Similar to the system used by Knoch and Speck~\cite{knoch2017nonequilibrium}, and Alexander-Katz et al. ~\cite{alexander2006shear}, we consider a polymer containing 50 monomers evolving with an underdamped Langevin equation
\begin{align}\label{eq:5.3}
    m \mathbf{\dot v}_i &= -\gamma \mathbf{v}_i - \nabla_{\mathbf{r}_i} V(\mathbf{r}^N) + \mathbf{F}^a_i(\mathbf{r}^N)  + \boldsymbol{\eta}_i  \\
    \mathbf{\dot r}_i &= \mathbf{v_i} \notag
\end{align}
where $\mathbf{r}_i$ and $\mathbf{v}_i$ are the position and velocity vectors of the $i$th monomer, $\gamma$ is the friction coefficient, and $\boldsymbol{\eta}_i$ is Gaussian random noise with mean $\langle \boldsymbol{\eta}_i(t) \rangle=0$ and variance $\langle \boldsymbol{\eta}_i(t) \otimes \boldsymbol{\eta}_j(t) \rangle=2 \gamma k_{\mathrm{B}}T\delta_{ij}\delta(t-t')\mathbf{I}_3$. The conservative potential $V(\mathbf{r^N})$ takes the form
\begin{equation}\label{eq:5.4}
    V(\mathbf{r}^N) = V^\mathrm{LJ}(\mathbf{r}^N) + V^{\mathrm{bond}}(\mathbf{r}^N) + V^{\mathrm{wall}}(\mathbf{r}^N)
\end{equation}\label{eq:5.5}
where $V^\mathrm{LJ}(\mathbf{r}^N)$ is a Lennard-Jones potential,
\begin{equation}\label{eq:5.6}
    V^{\mathrm{LJ}}(\mathbf{r}^{N}) = \sum_{i<j}^{N} \epsilon \left [ \left(\frac{r_{ij}}{\sigma}\right)^{12} - 2 \left(\frac{r_{ij}}{\sigma}\right)^{6}\right ]
\end{equation}\label{eq:5.7}
$V^{\mathrm{bond}}(\mathbf{r}^N)$ is the harmonic bond potential between adjacent beads,
\begin{equation}
    V^{\mathrm{bond}}(\mathbf{r^N}) = \sum_i^{N-1} \frac{\kappa}{2} \left(\frac{r_{i,i+1}}{\sigma} -1\right)^2
\end{equation} with $r_{i,i+1}$ denoting the distance between beads $i+1$ and $i$, and $\kappa$ is the spring constant. The polymer is placed in a periodic box of size $200\times200\times200 \sigma^3$ and grafted onto the $x$-$y$ plane by fixing the location of the first bead to $(0.0,0.0,100.5\sigma)$ and adding a repulsive wall potential of the form $V^{\mathrm{wall}} (\mathbf{r}^N) = \epsilon_{\mathrm{wall}} \sum_{i}^N z_i^{-12}$. Finally, the active force is given by a simple shear flow of the form
\begin{equation}\label{eq:5.8}
    \mathbf{F}^{\mathbf{a}}_i = \dot \gamma (z_i-z_c) \hat{\bf{x}}
\end{equation}
where $\dot \gamma$ is the strain rate, $z_c$ is the $z^{\mathrm{th}}$ component of the center of mass of the polymer, and $\hat{\bf{x}}_i$ is the unit vector in the $x$ direction. %This form of the flow has been used previously to illustrate the applications of nonequilibrium Markov state modelling, and is shown to exhibit the cyclic dynamics observed experimentally in vWF proteins.~\cite{knoch2017nonequilibrium} 
For the simulations parameters, we use $\sigma=m=1,\;\kbt=2.0,\;\epsilon=4.6,\; \kappa=500.0, \;\epsilon_{\mathrm{wall}}=4.0, \; \gamma=10.0$, $\beta \dot \gamma \sigma^2=6.5$, reduced time units $\tau = \sqrt{2m\sigma^2/\kbt}=1$ and timestep $dt=10^{-2}\tau$. All simulations reported in this section are performed in OpenMM~\cite{eastman2017openmm}, and the Normal Mode Langevin Integrator~\cite{sweet2008normal} is used to discretize the equations of motion. 

For the specific strain rate studied $\dot \gamma=10$, we find the polymer to exhibit transitions between a stretched and collapsed states visualized in Fig. \ref{fig:9} (A). The time-series of the end-to-end vector ${\color{black}R_N=|\mathbf{r}_N - \mathbf{r}_1|/N\sigma}$ for a long simulation for the finite shear case is plotted in Fig. \ref{fig:9} (B) and shows that the two states are long-lived with relatively long transition path times. The extended state is highly thermodynamically unfavorable in equilibrium, as shown by the instantaneous relaxation of the extended state of the polymer to the globular state, plotted in red in Fig. \ref{fig:9} (B). The instability of the extended state in equilibrium and emergent metastability under finite shear is also evident from the plot of the negative log-probability along the end-to-end distance, $\ln p(R_N)=\ln \langle \delta[R_N-R_N(t) ]\rangle$, as shown in Fig. \ref{fig:9} (C). For the specific value of shear, a barrier of height approximately $1 \kbt$ relative to the globular state and $0.25 \kbt$ relative to the stretched state is observed in the figure.

Similar to the previous section, we use a variational method~\cite{singh2023variational} to compute the time-dependent committor from the reactive trajectory ensemble to gain insight in the dynamics. Specifically for this system, we are interested in obtaining the dominant modes that are responsible for the transition between the two metastable states that emerge out of equilibrium. There are two aspects that make inference into this system specifically challenging. First, there is a very small gap within the spectrum of the Fokker-Planck generator, since the relaxation timescales of the polymer compete with the timescales of transition between the globular and the stretched state. While this makes it considerably easier to obtain a reactive trajectory ensemble, the reactive trajectories themselves are longer. Second, there are no good set of descriptors available \emph{a-priori} since the bond-angle-transformation generally employed for representing chain molecules~\cite{singh2023variational} do not form a complete set of descriptors due to the anisotropy of the system induced by the shear force, as well as the repulsive planar wall at $z=0$. %Hence, for a lack of a better choice, we use the raw coordinates to perform optimization using our method. 

Besides the two reactive processes that occur at finite shear, we additionally consider the equilibrium excitation process to understand the dominant modes the fluctuations would take place if the reaction took place in equilibrium. To perform the same analysis for equilibrium, we chose random configurations from the nonequilibrium steady-state distribution within the unfolded state. From those states, we run simulations of trajectories for $t_f=400\tau$, in which the polymer is observed to relax to globular state with a probability close to 1. These trajectories are then time-reversed, and the same optimization is performed on those trajectories. Details of the optimization and the ansatz are in Appendix \ref{appendD}. 

In order to unpack the reaction mechanisms, we consider the bond vector transformation where $\mathbf{\tilde r_j}$ is given by
\begin{align}\label{eq:5.9}
    \mathbf{\tilde r}_j = \begin{cases}
    \mathbf{r}_1,& j= 1\\
    \mathbf{r}_{j+1} - \mathbf{r}_j,  & \text{otherwise}
\end{cases}
\end{align}
where we note that the first monomer denoted by $\mathbf{r}_1$ is not included in the transformation as its position is constrained due to grafting.
The Jacobian of the transformation $\mathbf{r} \to \mathbf{\tilde r}$ is a $3 N \times 3 N$ matrix given by
\begin{align}\label{eq:5.10}
    J_{ij}  = \begin{cases}
    1,& i= j\\
    -1,  & i= j-3 \\
    0, & \text{otherwise}
\end{cases}
\end{align}
This is a complete but nonorthogonal coordinate transformation, where the diffusion weighted metric tensor~\cite{singh2023variational} $\boldsymbol{\Gamma}^{-1}=\mathbf{J}^T \mathbf{J}$ is given by
\begin{align}\label{eq:5.11}
    \Gamma_{ij}  = \begin{cases}
    1,& i=j\leq d \\
    2,& i=j\geq d \\ 
    -1,& i=j\pm d \\ 
    0, & \text{otherwise}
\end{cases}
\end{align}
which implies that $\boldsymbol{\Gamma}^{-1}$ is a special case of tridiagonal, symmetric, Toeplitz-like matrix. By diagonalizing this tensor using an eigenvalue decomposition, one obtains the normal modes similar to Rouse modes, $\mathbf{\bar r^N}$, where the $i$th eigenvalue $\epsilon_i$  and the $j$th component of the $i$th eigenvector $\nu_{ij}$ is given by~\cite{yueh2005eigenvalues}
\begin{align}\label{eq:5.12}
    \epsilon_i &= 2 \left( 1-\cos\left( \frac{\pi (2i-1)}{2N-1}\right) \right) \notag \\ 
    \nu_{ij} &= n_{i} \cos\left( \frac{\pi(2i-1)(2j-1)}{2(2N+1)} \right)
\end{align}
where $n_{i}$ is the normalization constant, which in the large $N$ limit is $n_j=1/\sqrt{N/2}$.
In this case, the force along the modes $\boldsymbol{\bar \lambda}$ is given by
\begin{equation}\label{eq:5.13}
    \bar \lambda_i = \frac{n_i}{\epsilon_i} \sum_j  \lambda_j  \sin\left(\frac{\pi(2i+1)j}{2N+1}\right)
\end{equation}
where $\frac{n_i}{\epsilon_i} \sin\left(\frac{\pi(2i+1)j}{2N+1}\right) $ is the $ij$ element of the inverse of the Jacobian matrix $ \mathbf{\bar J}$ of the $\mathbf{r} \to \mathbf{\bar r}$ transformation, given by $ \mathbf{\bar J} = \mathbf{J} \boldsymbol{\nu_0}$.

\begin{figure}[t]
  \centering
    \includegraphics[width=8.2cm]{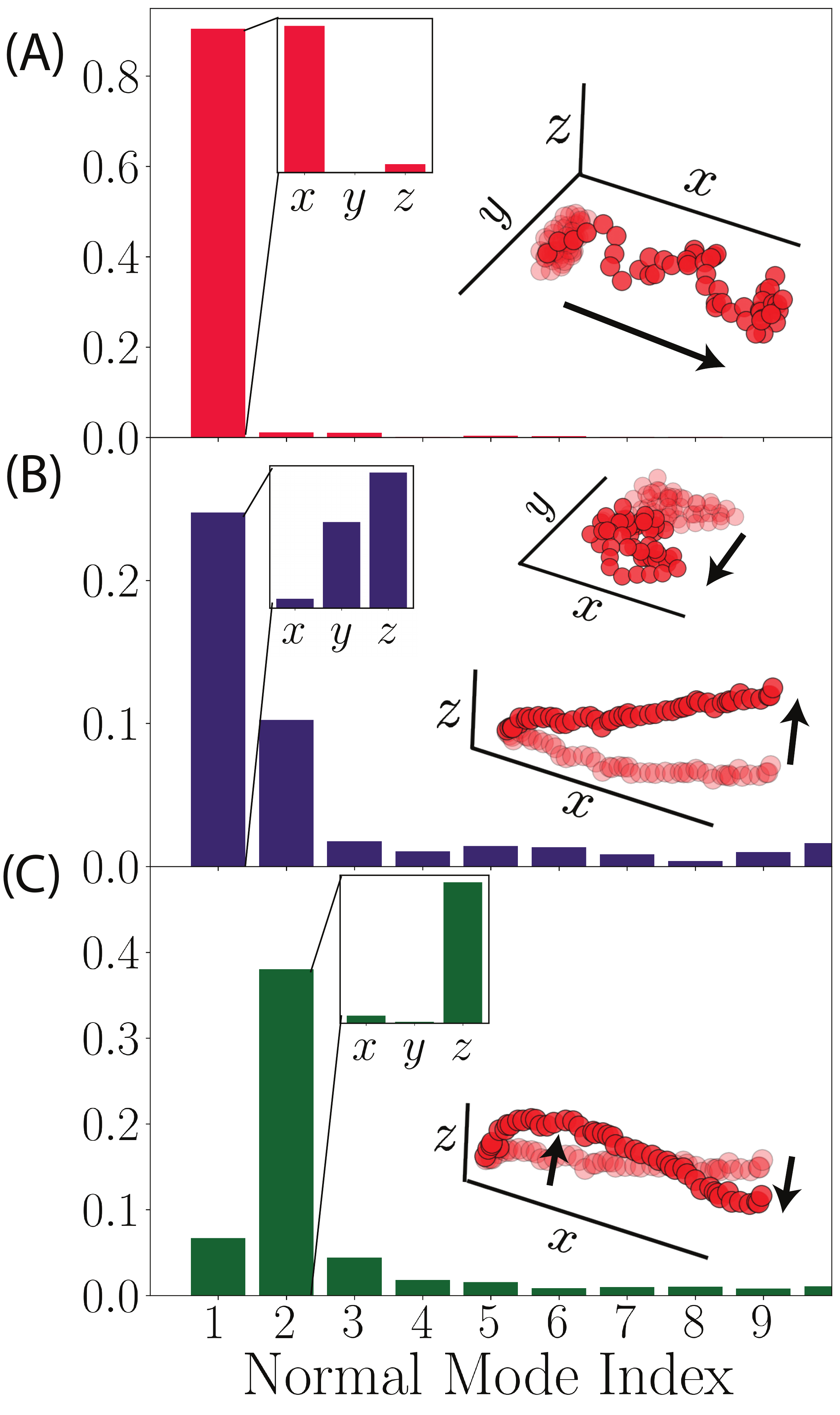}
    \caption{Dominant modes that are activated during the three different processes. Relative action decomposition for the equilibrium excitation (A), nonequilibrium unfolding (B) and nonequilibrium folding (C) in terms of the normal modes of the polymer. While the first two modes are observed to be activated for all three processes, substantial mechanistic differences are observed between the unfolding and folding pathways, all of which are also different from the excitation event in equilibrium. These differences can be distilled by considering the per component decomposition of the dominant modes shown within the left inset of each figure. A visualization of the dominant modes are shown in the right insets.}
    \label{fig:10}
\end{figure}

\begin{figure*}[t]
  \centering   \includegraphics[width=0.99\textwidth]{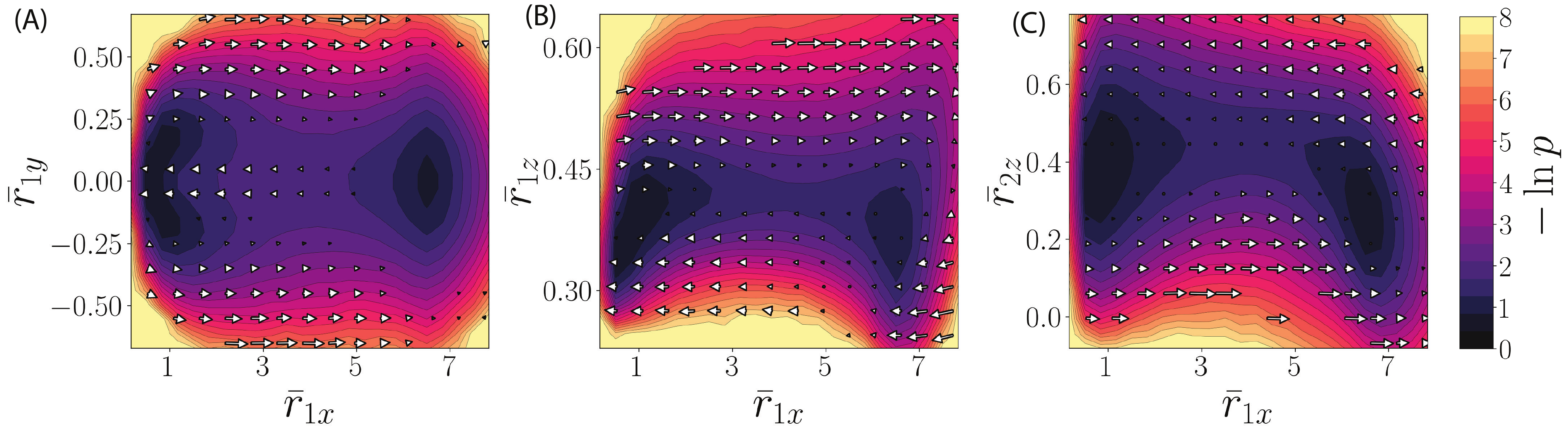}
    \caption{Understanding the mechanism of the three different processes. (A) - (C) Marginalized log probability density and the Fokker-Planck velocity along the leading normal modes identified to be activated during the equilibrium and nonequilibrium transitions.}
    \label{fig:11}
 \end{figure*}
We compute the contribution to the rate of each process by computing the transformed forces acting on the normal modes from the originally optimized control force using Eq. \ref{eq:5.13}, and then computing the stochastic action using the analogs of Eq. \ref{eq:decomp} for the underdamped integrator. These contributions for the equilibrium process as well as the nonequilibrium unfolding and folding transitions are shown in Fig.~\ref{fig:10} (A), (B) and (C), respectively. We find that all three processes are dominated by the first 2 normal modes of the polymer. For the equilibrium relaxation process, the first mode completely determines the relaxation process and accounts for approximately 90\% of the total action. The first mode is found to be important for the nonequilibrium unfolding transition, however the contributions are delocalized among other modes. This delocalization is observed in the nonequilibrium folding process as well, however the transition is strongly determined by the second mode. 

We next consider the individual components of the normal modes to gain further insight. The decompositions along the components of the normal modes are shown in the insets within Fig. \ref{fig:10} (A), (B) and (C) for the equilibrium excitation and nonequilibrium unfolding, folding process respectively reflecting the anisotropy induced by the repulsive wall and the shear force. All three processes are well described by different descriptors, clearly illustrating the modes activated for the three processes are different from each other. The equilibrium relaxation process is well described by the $x$ component of the first mode $\bar r_{1x}$, also visualized in the inset of Fig. \ref{fig:10} (A). The mode $\bar r_{1x}$ corresponds to the stretching of the polymer along the $x$ axis, and is straightforward to understand by considering the form of the system. The monomers in the system are strongly attractive through the Lennard-Jones potential, and the relaxation of the polymer in equilibrium closely resembles the decay towards the minima of an excited particle in a harmonic potential ~\cite{rubinstein2003polymer}. The $\bar r_{1x}$ mode essentially contains the harmonic free energy surface that transforms at some finite shear rate. 

Under the applied shear, we find that $\bar r_{1x}$ is minimally activated during the nonequilibrium reactions. This is counter-intuitive since the size of the shear is such that elongating the end-to-end distance is still energetically unfavorable. Instead, we find the $\bar r_{1y}$ and $\bar r_{1z}$ to become activated during the unfolding reaction and $\bar r_{2z}$ to become activated for the folding reaction. All of these modes are visualized within the insets of Figs. \ref{fig:10} (B) and (C). To fold in a shear flow, the polymer must obtain a rare fluctuation in the direction of the shear gradient that acts to buckle the chain.  To unfold in a shear flow, the polymer must obtain a rare fluctuations in the direction of the shear gradient as well as orthogonal to it. From this analysis it is clear that the mechanism for folding and unfolding are distinct out of equilibrium. 

To gain further insight into the two transitions within the nonequilibrium steady state, we compute the marginalized log probability and the net probability flux along the three dominant modes from a long trajectory. The probability is defined as,
\begin{equation}
\ln p(\bar{r}_i,\bar{r}_j) = \ln \langle \delta[\bar{r}_i -\bar{r}_i(t)] \delta[\bar{r}_j -\bar{r}_j(t)] \rangle
\end{equation}
where the average is over the nonequilibrium steady state. The plots for the log-probability along $\bar r_{1x}$ and $\bar r_{1y}$, along $\bar r_{1x}$ and $\bar r_{1z}$ and along $\bar r_{1x}$ and $\bar r_{2z}$ are shown in Figs. \ref{fig:11} (A), (B) and (C), respectively. Also shown is the marginalized Fokker-Planck velocity field~\cite{van2010three} defined as
\begin{equation}
\mathbf{v}(\bar{r}_i,\bar{r}_j) =  \frac{\boldsymbol{\epsilon}}{\gamma}\langle  \mathbf{F} \rangle_{\bar{r}_i,\bar{r}_j} - \frac{\kbt \boldsymbol{\epsilon} }{\gamma}\nabla \ln p(\bar{r}_i,\bar{r}_j)
\end{equation}
where $\boldsymbol{\epsilon}$ is a diagonal matrix with values $\epsilon_i$, $\langle \mathbf{F} \rangle_{\bar{r}_i,\bar{r}_j}$ is the mean force conditioned on two of the normal modes. We estimate $\bf{v}(\bar{r}_i,\bar{r}_j)$ on a grid by counting the number of exits from its boundaries. The left well with values of $\bar r_{1x} < 2$ corresponds to the folded or the globular state, and the right well with $\bar r_{1x} > 6$ corresponds to the unfolded or the extended state. We also observe that the barrier exists purely along the $\bar r_{1x}$ component, consistent with our previous hypothesis on the equilibrium excitation process. 

The importance of each mode can be explained by considering the direction and the magnitudes of the probability currents. In Figs. \ref{fig:11}, we observe strong persistent probability currents between the two metastable states that generally increase in magnitude towards the extremal points along modes $\bar r_{1y}$, $\bar r_{1z}$ and $\bar r_{2z}$. In Fig. \ref{fig:11} (A), strong unidirectional currents are observed from the folded to the unfolded state at all values of $\bar r_{1x}$, and large absolute values of $\bar r_{1y}$. The direction of the current flips towards the folded state at small values of $\bar r_{1y}$. For Fig. \ref{fig:11} (B), strong currents towards the extended state are observed above the line $\bar r_{1z}=0.4$, and increase in magnitude with increasing value of $\bar r_{1z}$. Within the two metastable basins and below $\bar r_{1z}=0.4$, the currents are observed to flip directions and increase in magnitude with decreasing $\bar r_{1z}$. Due to the strong repulsive interactions at intermediate values of $\bar r_{1x}$ and values of $\bar r_{1z}$ below 0.4, no currents are observed within the system. The trends in Fig. \ref{fig:11} (C) are quite similar to that of Fig. \ref{fig:11} (B), where strong currents are towards the unfolded state are observed below the line $\bar r_{2z} \approx 0.4$, above which the direction of the currents switch towards the folded state.  

Given this, the mechanism of the two reactions out-of-equilibrium can be understood as follows. The important  event in unfolding is the activation of $\bar r_{1z}$, which corresponds to stretching of the polymer along the $z$ axis, as shown in the inset of Fig. \ref{fig:10} (B). 
Persistent currents towards $+\bar r_{1x}$ are observed with increasing values of $\bar r_{1z}$, and the currents switch direction about a line of symmetry and act along the $-\bar r_{1x}$ with magnitude increasing away from the line of symmetry. Hence, to unfold, the mechanism involves extending along $\bar r_{1z}$, against the minimum free energy path, and coupling to these persistent probability currents that carry it to the unfolded state. The importance of the $\bar r_{1y}$ mode for the unfolding reaction is interesting because $y$ is the transverse direction that is not directly coupled to the shear force. Based on  Fig. \ref{fig:11} (A), we find that the minimum free energy path that lies along the line $\bar r_{1y}=0$ contains strong unidirectional currents that point towards the folded state. These results can be attributed to the steric interactions of the polymer under the effect of shear, which in turn causes the uncoupled mode to become activated for the reactive process, making it more probable to fold for small values of $\bar r_{1y}$, and more probable to unfold at large values of $\bar r_{1y}$. In order to react, the polymer requires additional fluctuations towards the extremal values of $\bar r_{1y}$, in order to couple to the currents that take it to the unfolded state.

To understand the mechanism of the folding reaction, we reconsider the plot along $\bar r_{1x}$ and $\bar r_{1z}$. The wall induces repulsive interactions that prohibit the polymer from stretching down along the $z$ axis, strongly destabilizing the region where the currents carry the polymer to the folded state. Hence, to fold, the polymer has to instead fluctuate along the buckling mode $r_{2z}$, that is visualized in the right inset in Fig. \ref{fig:10} (C). The mechanism is similar to the that of unfolding, in that the reaction involves coupling to strong currents at the top of $\bar r_{2z}$ that arise due to action of the flow field with the latter half of the polymer. The distinct pathways that transition the system to fold or unfold in the presence of the shear afford an opportunity to accumulate probability in the unfolded state, stabilizing it by kinetic trapping even though it is energetically unstable.

\subsection{When does a breakdown in microscopic reversibility generate stability?}

When non-conservative forces are added to a system, the pathways that transition a system between two states need not be the same in the forward and backwards direction. This is a reflection of the existence of persistent probability currents, stabilized by a finite rate of energy injection. Steady-state probability densities under such nonequilibrium conditions depend on the rates at which probability flux enter and leave a region of phase space. Kinetic factors irrelevant in a system for which detailed balance holds, can retard probability flux out of a region, allowing for the accumulation of probability and the generation of metastability. These kinetic factors can arise because of entropic or geometric constraints that render motion in some direction slow, or because barriers between pathways of some reactive channels are larger than other. We have shown that when nonequilibrium driving forces compete against these slow dynamics, they can tip the balance of probability fluxes. Clearly the harnessing of probability fluxes relate to the generation of entropy, and the connection between the stabilization of states out of equilibrium found here and the Kawaski-Crooks relationship that relates equilibrium and nonequilibrium distribution functions is worthy of clarification~\cite{rosa2024variational}.

As a example, we have considered the shear induced extension of a polymer. Such systems have been widely studied for a experimentally and theoretically ~\cite{de1974coil,smith1999single, perkins1995stretching, perkins1997single, jaspe2006protein, huang2010semidilute, doyle2000dynamics, delgado2006cyclic,zhang2009tethered,lumley1969drag}. Of relevance to this paper is the early results based on rheological arguments~\cite{de1974coil} that found that a simple shear flow by itself to be inadequate in completely stabilizing the strongly stretched state of semiflexible polymers in dilute solution, because the rotational component of the shear flow induced an end-to-end tumble motion that destabilized the extended state. Experimentally these results were verified by studying a single DNA under different flow fields ~\cite{smith1999single,perkins1995stretching,perkins1997single}, where it was found that compared to an elongation flow field, the extension of a DNA was highly suppressed under a shear flow. Additionally for the case of shear flow, the DNA was observed to display cyclic motion that were reported in computational studies. Subsequently, it was reported that by acting a shear flow on a tethered polymer~\cite{doyle2000dynamics}, the stability of the polymer was substantially enhanced, although the polymer was still observed to exhibit cyclic dynamics, and exhibit fluctuations between the globular and the extended state. 
%Within this context of our study, it is intuitive that the repulsive wall plays a strong role in redirecting the currents back towards the extended state, and suppresses reactions back to the globular state. 

Within the context of these studies, we reconsider the log probability density along the $\bar r_{1x}$ and $\bar r_{1z}$ mode in Fig. \ref{fig:11} (B), specifically focusing on the repulsive interactions at $2 < \bar r_{1x} < 5$ and $0.2 < \bar r_{1z} <0.45$. As mentioned earlier, the equilibrium free energy of the polymer is symmetric about $ \bar r_{1x}$ and $ \bar r_{1y}$, but not along $ \bar r_{1z}$ due to the strongly energetic interactions between the polymer and the repulsive wall. 
%Fig. 11 (A) additionally highlights the existence of a strong repulsive region that fades away with increasing values of $\bar r_{1x}$. 
The geometric nature of the repulsive interactions are due to entropic effects that make it significantly easier for the polymer to stretch down towards the repulsive wall when it is stretched along $\bar r_{1x}$, compared to when the polymer is in its globular state or half-extended state. This is attributed to the polymer being able to buckle up or move along other orthogonal $z$ modes to decrease contact with the wall. The effect of the adding shear along the two stretching modes leads to generation of persistent probability currents along $\bar r_{1x}$ that switch signs about $\bar r_{1z}$ = 0.4. The currents on the top of $\bar r_{1z}$ increases the flux towards the extended state, but are not sufficient in completely stabilizing the extended state~\cite{doyle2000dynamics}. The key factor appears to be the interaction of the incoming current at bottom of the plot with repulsive wall. This description strikes a strong similarity with the two-dimensional system and the three site Markov state model discussed in the previous subsection. Based on these observations, the polymer has all the necessary ingredients as the low-dimensional system and the Markov state model model discussed in the previous subsection -- an increased flux towards a thermodynamically unstable state induced by nonequilibrium driving coupled with strong repulsive interactions due to the wall that lead to asymmetry in the flux balance.

\begin{figure}[t]
    \centering
    \includegraphics[width=0.9\linewidth]{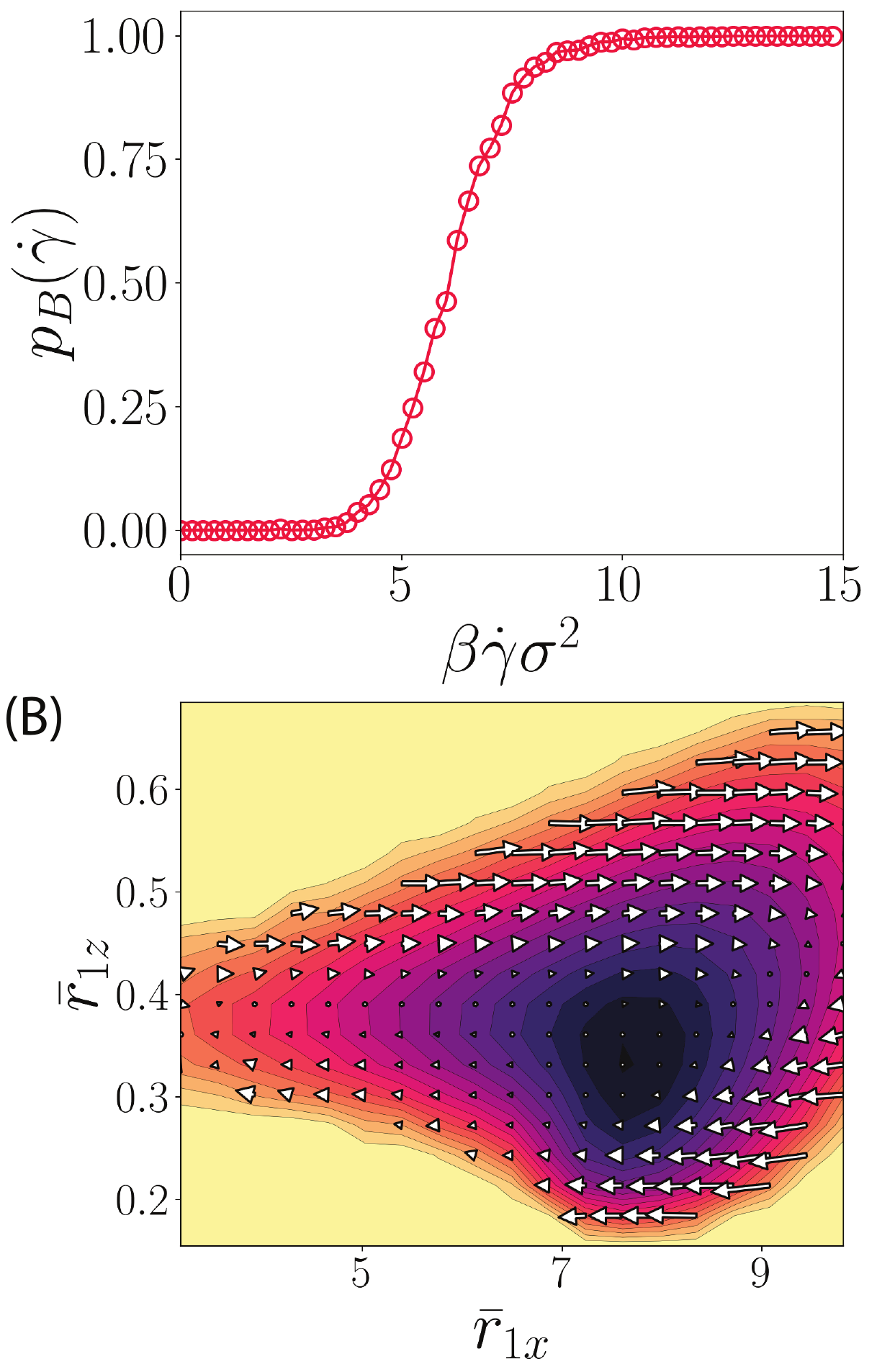}
    \caption{(A) Probability enhancement of the unfolded state as a function of strain rate. (B) Marginalized log probability density and the Fokker-Planck velocity field along modes $\bar r_{1x}$ and $\bar r_{1z}$ for $\beta \dot \gamma \sigma^2= 15$.}
    \label{fig:12}
\end{figure}

To further explore this similarity, we consider the enhancement in the probability of the unfolded state $p_B$ as a function the strain rate:
\begin{equation}
    p_B(\dot \gamma) = 1 - \langle \Theta(R_N - 0.2 ) \rangle
\end{equation}
where $\langle \cdots \rangle$ denotes steady-state averages computed from long trajectories at a specific strain rate $\dot \gamma$. The specific choice of defining indicator functions is done to avoid choosing strict definitions of the emergent state. The plot of the probability enhancement is shown in Fig. \ref{fig:12} (A). The plot shows a strong qualitative agreement with the enhancement of the two systems discussed in Fig. \ref{fig:8} (A). Additionally, we consider the log probability density and marginalized Fokker-Planck velocity field along the $\bar r_{1x}$ and $\bar r_{1z}$ modes for the highest strain rate considered corresponding to $\beta \dot \gamma \sigma^2= 15$, shown in Fig. \ref{fig:12} (B). Similar to Fig. \ref{fig:8} (B), we find strong cyclic currents in the clockwise direction along the region corresponding to the extended state. Strong currents towards the right well above the line of symmetry along  $\bar r_{1z}$ that eventually hit the repulsive interactions at large values of $\bar r_{1z}$ and $\bar r_{1x}$, that corresponds to the polymer being stretched completely. These currents get redirected downwards, and are observed to act towards the left well corresponding to the extended state. Due to the nonlinear interaction located below the symmetry line along $\bar r_{1z}$ and intermediate values of $\bar r_{1x}$, these currents are redirected up and restart the cycle again. The existence of these cycles also corroborate the experimental and computational results on the cyclic motion of the last monomer along the $x-z$ plane, closely connected to the tumbling of the polymer under shear flow \cite{zhang2009tethered,doyle2000dynamics,schroeder2005characteristic,gerashchenko2006statistics}.
Within this context, it is intuitive to understand why addition of shear flow to free polymers, i.e. untethered polymers cannot completely stabilize the extended state.\cite{de1974coil,smith1999single, doyle2000dynamics} The repulsive wall plays a strong role in redirecting the currents back towards the extended state, and suppresses reactions back to the globular state. Finally, we believe that the results of stability arising from interactions between repulsive interactions and steady currents are also closely related to the accumulation of active particle under confinement and against hard surfaces~\cite{nikola2016active,fily2014dynamics,fily2015dynamics,smallenburg2015swim,mallory2014anomalous}.

\section*{Concluding discussion}
In this work, we have highlighted how kinetics and dynamics of systems in equilibrium are perturbed by addition of nonequilibrium forces. Our approach was to actively consider the importance of probability currents, breakdown of detailed balance and the existence of more than one reactive pathways. This was made possible by dynamically-consistent coarse graining of high dimensional interacting systems evolving within nonequilibrium steady states. Advances in variational path sampling, dynamic control and trajectory reweighting enabled the discovery of such a low dimensional representations without invoking intuitions from equilibrium systems. Through this approach, we were able to elucidate mechanisms of complex, driven systems and uncover some basic principles that dictate their rare, reactive events in ways distinct from near equilibrium ideas. 

\section*{Acknowledgements}
We would like to thank Dr. Avishek Das and Dr. Jorge Rosa-Raíces for invaluable discussions regarding variational path sampling methods, and Sam Oaks-Leaf for illuminating discussions regarding dynamical coarse graining of the systems in this work using Markov state models. This work was supported by the U.S. Department of Energy, Office of Science, Office of Advanced Scientific Computing Research, and Office of Basic Energy Sciences, via the Scientific Discovery through Advanced Computing (SciDAC) program.

\section{Appendix}
\subsection{Optimization of the committor}\label{appendA}
We use the variational method based on stochastic optimal control~\cite{singh2023variational} to compute the committor. 
Using the Jacobi-Bellman-Hamilton~\cite{das2022direct} equation and spectral theory~\cite{singh2024splitting}, the optimal force can be approximated by the steady-state committor $q_B(\mathbf{r})$ as given in Eq. \ref{eq:doob}. Under the separation of timescales, the optimal force for the backward reaction $B \rightarrow A$ is given by
\begin{equation}
    \lambda^*_{A}(\mathbf{r},t_f-t) \approx -\lambda^*(\mathbf{r},t_f-t)\frac{q_B(\mathbf{r}) + k_{AB}(t_f-t)}{1-q_B(\mathbf{r}) - k_{AB}(t_f-t)} \, .
\end{equation}
This follows from expressing the time-dependent committor $q_B(\mathbf{r},t)$ for the backward process in terms of the steady-state-committor $\bar q_A(\mathbf{r})$ and the steady state probability density $\bar p_A$ and $\bar p_B$ for the $A$ and $B$ states respectively~\cite{singh2024splitting}
\begin{align}\label{eq:s6}
   q_A(\mathbf{r},t_f) &= \bar q_A (\mathbf{r}) e^{-\mu_2 (t_f-t) } + \bar p_A (1-e^{-\mu_2 (t_f-t)}) \notag \\
%   &= (1-q_B(\mathbf{r})) e^{-\mu_2 (t_f-t) } + \bar p_A (1-e^{-\mu_2 (t_f-t) }) \notag \\
%   & \approx  1-q_B(\mathbf{r})e^{-\mu_2 (t_f-t) } - \bar p_B (1-e^{-\mu_2 (t_f-t) }) \notag \\
%   & = 1-q_B(\mathbf{r},t_f-t) \notag \\
   & \approx 1-\bar q_B(\mathbf{r})-k_{AB}(t_f-t)
\end{align}
where we have used $\bar q_A(\mathbf{r}) = 1-\bar q_B(\mathbf{r})$ and assumed the system is bistable, so that $\bar p_B = 1- \bar p_A$. For systems in equilibrium where the reactive trajectory ensemble for the backward process can be obtained by time-reversing the reactive trajectory ensemble for the forward reaction, this form offers a simple and efficient way to optimize for the steady-state committor by simply choosing an appropriate ansatz for the optimization of the forces. 

We chose a time-independent ansatz for the steady-state committor $q_{B}(\mathbf{r};\boldsymbol{\theta})$ parameterized by parameters $\boldsymbol{\theta}$, and an additional parameter for time $\phi$ related to the rate, such that the time-dependent committor is \begin{equation}
    q_B(\mathbf{r},t_f-t;\boldsymbol{\theta},\phi) = \bar q_B(\mathbf{r};\boldsymbol{\theta}) + e^{\phi} (t_f-t) \, .
\end{equation}
The loss function $\mathcal{G} (\theta,\phi)$ can be given by the sum of the loss function for the forward and the backward process:
\begin{equation}
    \mathcal{G} (\boldsymbol{\theta},\phi) = \mathcal{G}_{AB} (\boldsymbol{\theta},\phi)+ \mathcal{G}_{BA}(\boldsymbol{\theta},\phi)
\end{equation}
and the forward loss function $\mathcal{G}_{AB} (\boldsymbol{\theta},\phi)$ expressed in terms of the time-dependent committor 
\begin{align}
     \mathcal{G}_{AB} (\boldsymbol{\theta},\phi) &= \bigg{\langle} \sum_{n=0}^{(t_f-1)/\Delta t} \sum_{i}^N \frac{\kbt \Delta t}{\gamma_i}\frac{(\partial_{r_i} q_{B}^2(\mathbf{r},t_f-n\Delta t)}{q_B^2(\mathbf{r},t_f-n\Delta t)} \notag\\
     &- \frac{\Delta t}{\gamma_i }\frac{(\partial_{r_i} q_{B}(\mathbf{r},t_f-n\Delta t)) \eta_i(n\Delta t) }{q_B(\mathbf{r},t_f-n\Delta t)} \bigg{\rangle}_{B|A}
\end{align}
where the first sum runs over all the timesteps within the reactive trajectory ensemble, and the second sum runs over all the system's degrees of freedom.  The noise along coordinate $i$ at step $n$, is calculated in the Ito sense $\eta_i (n\Delta t) = \gamma_i  (r_i((n+1)\Delta t)-r_i(n\Delta t)) / \Delta t - F_i[\mathbf{r}(n\Delta t)]$. For the backward process, the loss takes a similar form
\begin{align}
\mathcal{G}_{BA} (\boldsymbol{\theta},\phi) &=   \bigg{\langle} \sum_{n=\Delta t}^{\Delta t/t_f} \sum_{i}^N \frac{\kbt \Delta t}{\gamma_i}\frac{\partial_{r_i} q_{B}^2(\mathbf{r},n\Delta t)}{(1-q_B^2(\mathbf{r},n\Delta t))} \notag\\
    &+ \frac{\Delta t}{\gamma_i }\frac{(\partial_{r_i} q_{B}(\mathbf{r},n\Delta t))\tilde \eta_i(n\Delta t)}{1-q_B(\mathbf{r},n\Delta t)} \bigg{\rangle}_{B|A}
\end{align}
which corresponds to the discretized change in stochastic action for the time-reversed process written in the forward ensemble. For this case the discretized noise corresponds to that of the time-reversed process $\tilde \eta (n\Delta t) = (r_i((n-1)\Delta t) - r_i(n\Delta t))/\Delta t + F_i[\mathbf{r}(n \Delta t)]$. 

\begin{figure}[b]
  \centering
    \includegraphics[width=0.4\textwidth]{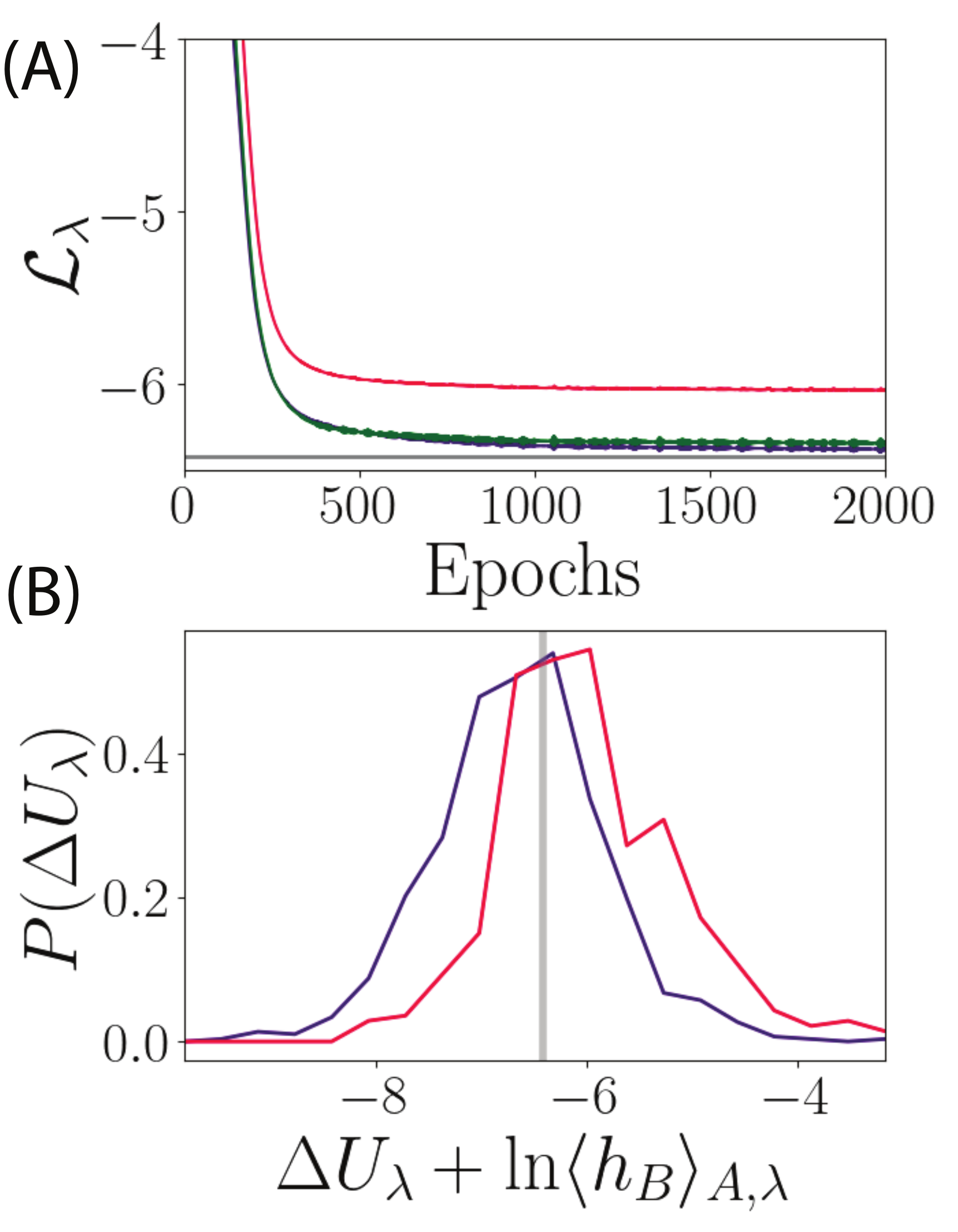}
    \caption{(A) Loss curves for the 2D system with a bistable potential given in Eq. \ref{eq:pes}. The colors denote first cumulant (red), second cumulant (purple), exponential estimator (green) along with the numerical estimate of $\ln kt_f$ (gray). (B) Shifted relative action distribution within the driven reactive (purple) and the original reactive ensembles (red).}
    \label{fig:a1}
\end{figure}
\begin{figure*}[t]
  \centering
    \includegraphics[width=0.99\textwidth]{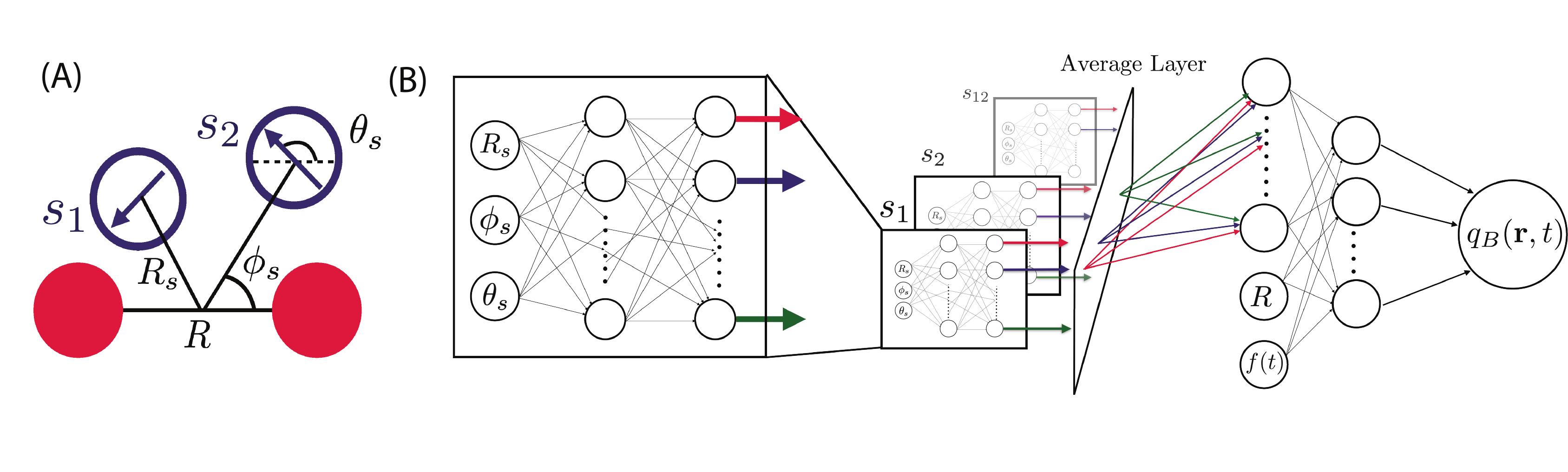}
    \caption{(A) Representation used for training the time-dependent committor for the passive dimer in active solvent  (B) Representative architecture of the neural-network ansatz used for training.}
    \label{fig:a2}
\end{figure*}
Additional Lagrange multipliers can be employed to constrain the values of the steady-dependent committor within wells $A$ and $B$, but are not necessarily required for optimization. For this problem, we obtain a reactive trajectory ensemble of 400 trajectories with $t_f=2\tau$, and perform optimization with a RMSprop optimizer for 2000 steps with a simple $20\times20\times 1$ neural network with CELU functions, and the output connect to a sigmoid activation function. The training loss curves for the optimization are shown Fig. \ref{fig:a1} (A) along with the exponential and second cumulant estimators and the numerical estimate of $\ln kt_f$.  

As an additional validation, we drive the trajectories with the control force and consider the action distribution~\cite{singh2024splitting} along the driven and undriven reactive trajectory ensembles. This allows us to check a symmetry between the likelihood of a change in action
\begin{equation}
P_\lambda(\Delta U_\lambda) = \langle \Delta U_{\lambda}-\Delta U_{\lambda}[\bf{X}(t_f)] \rangle_\lambda
\end{equation}
in the driven ensemble in the presence of $\lambda$ or its absence, $P_0(\Delta U_\lambda)$.~\cite{singh2023variational} This symmetry relates the two through
\begin{equation}\label{eq:sym}
\frac{P_\lambda(\Delta U_\lambda) }{P_0(\Delta U_\lambda) }=e^{\Delta U_\lambda - \ln k t_f /\langle h_B \rangle_{A,\lambda}}
\end{equation}
where $\langle h_B \rangle_{A,\lambda}$ estimate of the transition probability in the driven ensemble.
We check this relation by driving 1000 trajectories with the control force and computing $\Delta U_\lambda$ within those that react. With the specific ansatz, we find the reactivity of the driven process $\langle h_B \rangle_{A,\lambda}$ to be $\approx$ 0.85, and the action distribution within the driven and original reactive ensemble is observed to show a strong overlap indicating optimality. This can be observed in Fig. \ref{fig:a1} (B) that shows the action distribution, shifted with the $\ln \langle h\rangle_{A,\lambda}$ estimate. The first cumulant estimator within both ensembles is found to be $\approx 0.3$ away from estimate of $\ln kt_f$, that is corrected completely by a second cumulant approximation to the exponential estimator, Eq. ~\ref{eq:rate_OM}. For comparison, computation of the time-dependent committor~\cite{singh2023variational} without linearizing of the time-dependence was found to be with 0.2 from the empirical estimate.

\subsection{Dependence of $k$ on self-propulsion}\label{appendB}
For simplicity, we consider a bistable quartic potential $V(x)$ with with a saddle point centered at $x=0$ and two metastable wells with minimas at $x=x_A$ and $x=x_B$. For the system the rate $k$ can be computed using Kramers' theory~\cite{limmer2024statistical}
\begin{align}
k &= \frac{1}{\beta\gamma} \left[ \int_{-\infty}^{x_B} dx e^{-\beta V(x)}  \int_x ^{\infty} dx' e^{\beta V(x')}\right]^{-1} \notag\\
&\approx \frac{1}{\beta \gamma}e^{\beta \Delta V} \left[ \int_{-\infty}^{\infty} dx e^{-\beta \omega_A^2(x-x_A)^2/2}  \right]^{-1} \times \notag\\
& \qquad   \left[ \int_{-\infty}^{\infty} e^{-\beta \omega_0^2  x^2/2}dx  \right]^{-1}
\end{align}
where the saddle point approximation is made in the second line to decouple the two integrals and write it down in terms of the barrier height $\Delta V = V(0)-V(x_A)$ and the reduced harmonic frequencies $\omega_A$ and $\omega_0$ at the minima $x_A$ and the maxima $x=0$ respectively. Under the action of a linear bias potential $-\nu x$, that corresponds to the exact alignment of an active brownian particle with the reaction coordinate, one can use linear response theory to get a change in the rate as a function of the biasing parameter $\nu$. This can be done by noting that the probability within the A well takes the form
\begin{align}
& \int_{-\infty}^{\infty} dx e^{-\beta (\omega_A^2 (x-x_A) ^2/2+ \nu x )} = \frac{\sqrt{2\pi}}{\sqrt{\beta} \omega_A} e^{-\beta (\nu x_A - \nu^2/2 \omega_A^2)}
\end{align}
Similarly the second integral takes the form
\begin{align}
& \int_{-\infty}^{\infty} dx e^{-\beta( \omega_0^2 x^2/2+ \nu x) } =  \frac{\sqrt{2\pi}}{\sqrt{\beta} \omega_0} e^{\beta \nu^2 /2 \omega_0^2}
\end{align}
Substituting these results in for the rate, we have a simple form for the log ratio of the rate as a function of the linear bias $\nu$
\begin{align}
\ln  \frac{k (\nu) }{k (0)}  &= \beta\nu x_A - \frac{ \beta \nu^2 (\omega_A + \omega_0) }{2 \omega_A  \omega_0 } \approx  \beta l \nu 
\end{align}
where we note that the frequencies $\omega_A$ and $\omega_0$ correspond to the relaxation timescales of the system within the metastable well and over the barrier respectively. In terms of the spectrum of the Fokker-Planck generator, these frequencies correspond to eigenvalues of higher eigenmodes, and since the system is bistable, they are considerably larger than the second dominant eigenvalue of the system that is related to the rate of reaction $(1/\omega_A + 1/\omega_0)^{-1} \gg k$. For the case of the active brownian particle, the $\nu_0^2$ term can be safely neglected even at large Peclet under the limit $D_r < \omega_0+\omega_B $, due to the separation of timescale between the diffusion of the director and the faster relaxation processes of the passive system. Hence an effective log linear dependence of the rate on  a single parameter forms an excellent approximation. While additional quadratic corrections can be made, they are not explored here.

\subsection{Committor computation for the passive dimer in active bath}\label{appendC}

{\bf Ansatz and descriptors for representation.} As shown in Fig. \ref{fig:a2} the descriptors chosen to represent the commitor are the distance between the monomers of the passive dimer $R$, the distance between the closest solvent particles to the center of the mass of the dimer $R_s$, the angle of the solvent particles relative to the bond vector connecting the monomers of the dimer $\phi_s$, and the angle of the director relative the bond vector $\theta_s$. The closest 12 solvent particles $s_i$ are used for representation, with the index $i=1$ denoting the closest particle to the center of mass of the dimer for a given configuration. While these set of descriptors chosen are physically motivated, they do not obey permutation symmetry of the solvent particles. Hence, to enforce this symmetry, we chose a modified neural network architecture in which the weights are shared among all the solvent particles, similar to a transformer architecture.~\cite{boffi2024deep} Visualized in Fig. \ref{fig:a2} (B), the neural network is divided into two submodules, where the first submodule comprises 3 layers that only takes the solvent descriptors as inputs. The output from the final layer is a 20 dimensional vector that is averaged over all 12 solvent particles. These solvent averaged descriptors are concatenated with the bond distance $R$ and a modified time function $f(t) = 1/(t_f-t+0.01)$ and passed into the second module comprising one hidden layer. A sigmoid activation function is used in the final layer, and provides the model's estimate of the time-dependent committor $q_B(\mathbf{r},t)$. Except for the final layer, CELU/SiLU activation functions are to prevent the optimization from being hindered by weight decay.

\begin{figure}[t]
  \centering
    \includegraphics[width=0.45\textwidth]{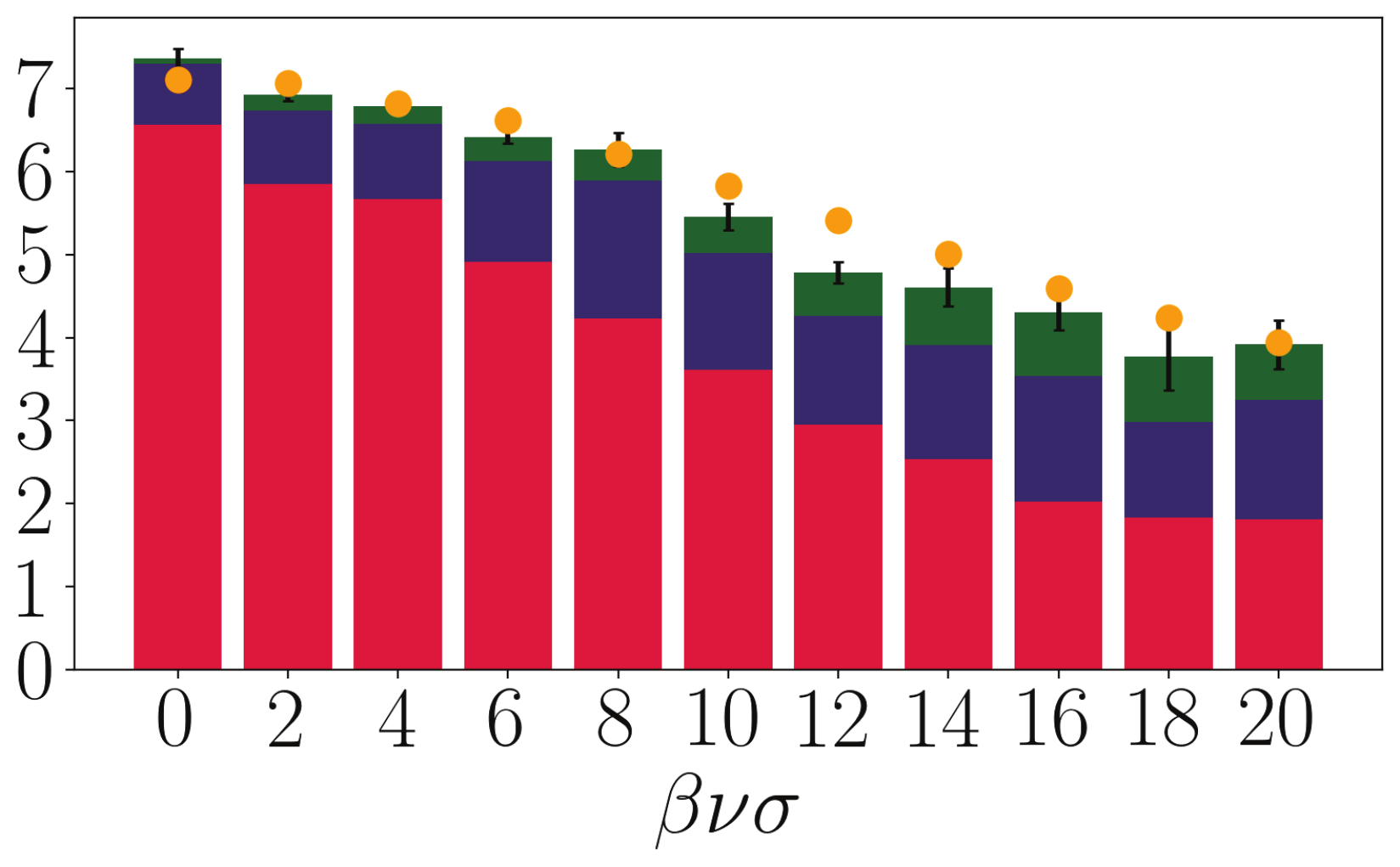}
    \caption{Final optimization results for the time-dependent committor for the dimer across different activity. The red and purple bars denote the first cumulant and the second cumulant estimates respectively, and the green bar denotes the additional correction due to $\langle \ln h_{B|A} \rangle_{\lambda}$ computed from driving 1000 reactive trajectories with the optimized control forces. The orange markers represent empirical estimates of $-\ln kt_f$.}
    \label{fig:a3}
\end{figure}
{\bf Optimization.} We obtain 300 reactive trajectories from brute force simulations for different Peclet in the range $\beta \nu \sigma = [0,2,4,6,8,10,12,14,16,18,20]$, and the transformed coordinates, noises and the Jacobian for the transformation are stored ahead of time. Due to the highly nonlinear nature of the solvent interactions, the first submodule of the ansatz that captures the solvent effect is optimized across all activity. This is done by optimizing the first submodule across all range of activity concurrently, while the second submodule is trained independently for each activity. Optimization is performed for 10000 steps using stochastic gradient descent, and after optimization, the first and the second cumulant estimates of the relative actions are found to be within values $0.05 - 0.5$ across the training and the validation dataset. The first and second cumulant estimates of the rate of are shown in Fig. \ref{fig:a3} along with the empirical estimate of $-\ln kt_f$ obtained from the brute force simulations. For small activity, it is found that the second cumulant estimator gives a robust estimate of $\ln kt_f$, but starts deviating as activity grows larger. By driving trajectories with the optimized controller for each activity, we find that the difference is due to not all driven trajectories being reactive. At $\beta \nu \sigma < 10$, approximately $75-95\%$ driven trajectories, which decreases to about $50\%$ for the highest activity considered. As shown in Fig. \ref{fig:a3}, the additional term of $\ln \langle h_B \rangle_{A,\lambda}$ is found to correct the estimator, and is found to be in excellent agreement with the rate estimate, within standard error.

{\bf Computation of $\Phi$.} To get an accurate and low variance estimate of the alignment of the director with the equilibrium controller, $\Phi(\nu)$, we retrain the time-dependent committor on the reactive trajectories in equilibrium. We decrease the number of activation functions per layer and only pass the distance $R_s$ and the relative angle of the solvent $\phi_s$ as inputs. We use the dual loss function based on the original and time-reversed reactive trajectories discussed in Appendix \ref{appendA} for optimization. Optimization is performed for 30000 steps using stochastic gradient descent. 

\begin{figure}[t]
  \centering
    \includegraphics[width=0.40\textwidth]{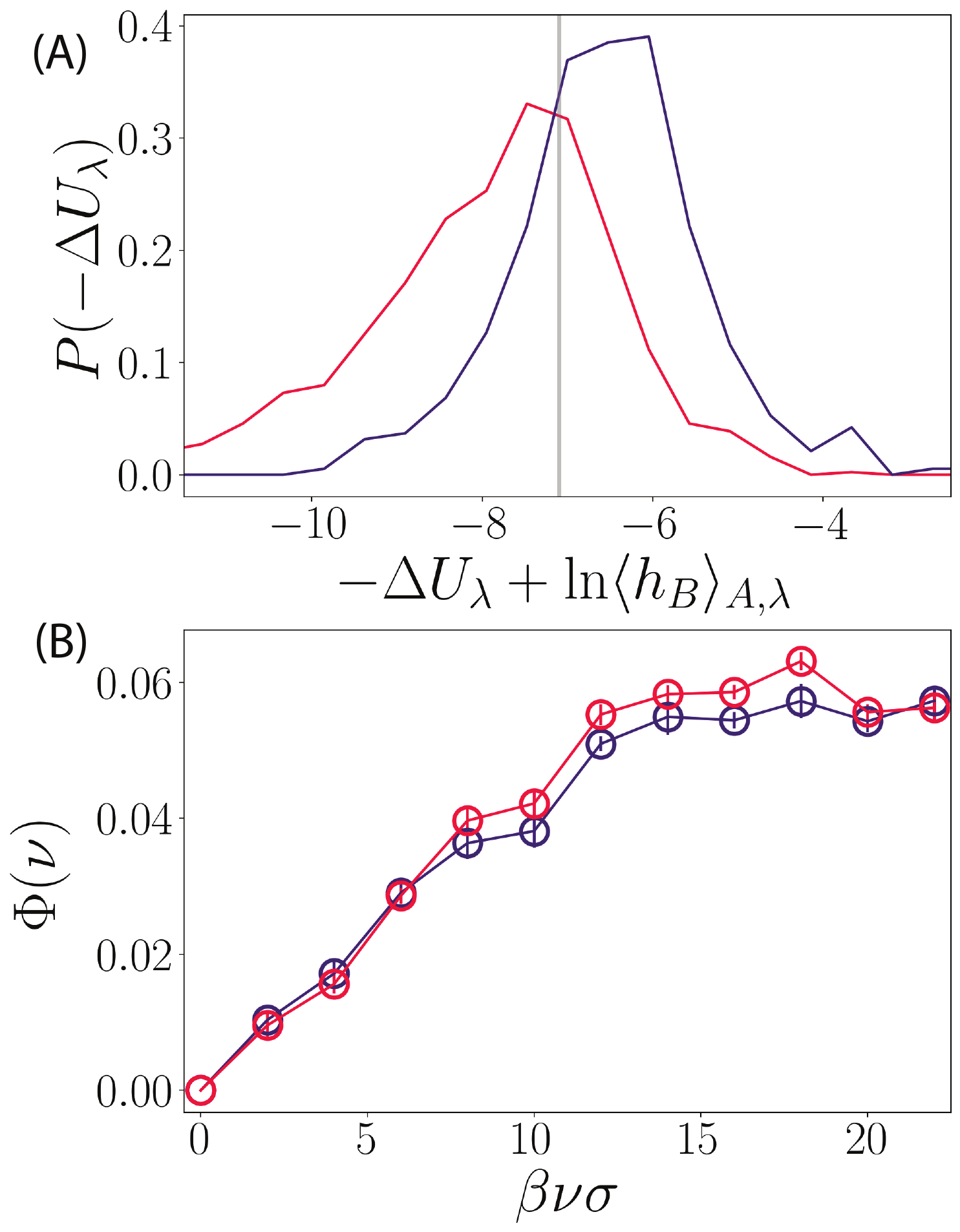}
    \caption{(A) Action distribution of the optimal controller for $\beta \nu \sigma=0$ in the driven (purple) and undriven (red) reactive trajectory ensemble along with the empirical estimate of $\ln kt_f$. The reactivity $\langle h_B \rangle_{A,\lambda}$ is 95.5\%. (B) Alignment of the director with the equilibrium committor that is computed using two different ansatze and optimization methods. Red denotes the same committor that is validated in (A), while purple labels the committor computed using the Behler-Parinello symmetry functions and optimized using a different variational loss function \cite{rotskoff2022active,khoo2019solving,hasyim2022supervised}.}
    \label{fig:a4}
\end{figure}

To validate the controller, we drive 1000 reactive trajectories from randomly drawn initial configurations within the collapsed state. From these runs, we find that 95.5\% of the trajectories are reactive, suggesting close to optimality. Additionally, we consider the distribution of stochastic action within the driven reactive trajectories and plot it in Fig.~\ref{fig:a4}, along with the distribution of the stochastic action in the undriven reactive trajectories. The strong overlap between the two distributions validates the optimality of the committor, from Eq. ~\ref{eq:sym}, barring numerical errors due to timestep discretization.~\cite{singh2024splitting} 

To evaluate $\Phi(\nu)$, we compute the normalized gradient of the equilibrium committor with the normalized director of the solvent particles that are closest to the center of the mass of the dimer, as defined in Eq. \ref{eq:phi_dimer}. Two modifications are made for better statistics. First, the averaging $\langle \hdots \rangle_{B|A}$ is performed in a Dirichlet sense, in that the averaging is only performed within sections of trajectories that last leave the collapsed state, defined through indicator function $R<1.25$, and first reach the stretched state defined by $R>1.85$. Second, given that only the 8 closest solvent particles are activated during the reaction at high activity, we truncate the sum $N_s = 8$ for better averaging. These modifications simply rescale the magnitude of $\Phi(\nu)$, but considerably decrease the standard error. The trend observed, of $\Phi(\nu)$ linear increasing until $\beta \nu \sigma = 10$ followed by subsequent plateau is the same regardless of these modifications.

We additionally chose a different ansatz and mode of optimization to further validate our results. Specifically, we use the Behler-Parinello symmetry functions implemented within the TorchANI along with the variational formalism \cite{rotskoff2022active,khoo2019solving,hasyim2022supervised} for solving the backward Kolmogorov equation. The form of the architecture is similar to the one used in our previous work \cite{singh2024splitting}, with the modification that the two monomers are defined as different species from the solvent particles. For the training dataset, 8 long trajectories of time $t=20000\tau$ with a time-lag of 0.2 $\tau$ are obtained. From these long trajectories, 150000 configurations are chosen within the regions inside $R<1.25$, $R>1.85$ and $1.25 \leq R \leq 1.85$. Optimization is performed for 10000 steps using stochastic gradient descent and post optimization, we compute $\phi(\nu)$ in the same way as before, by truncating the sum to $N_s=8$, and performing averaging over reactive trajectories (in the Dirichlet sense) for each activity. The results from these calculations are shown in Fig. \ref{fig:a4} (B) along with the results computed using the previously optimized controller shown in Fig. \ref{fig:6} (B). For all activity, the two independent estimates are found to be in excellent agreement in each other. 

\subsection{Leading descriptors for the grafted \\polymer under shear} \label{appendD}

All the optimization for the polymer were done with 300 trajectories as the training dataset and 50 trajectories as the validation set, obtained from brute force runs. Due to long relaxation times, a time-lag of 10 ($0.1 \tau$) was used within the configurations for the training set, and the noises were averaged over the $10$ steps. We found this form of averaging to considerably speed up convergence of an approximate controller, and found similar results for time lags up to 100 steps. The validation set comprised trajectories containing the positions and the noises over the whole trajectory, in order to obtain accurate metrics for the assessment of the accuracy of the controller.

\begin{figure}[b]
  \centering
    \includegraphics[width=0.4\textwidth]{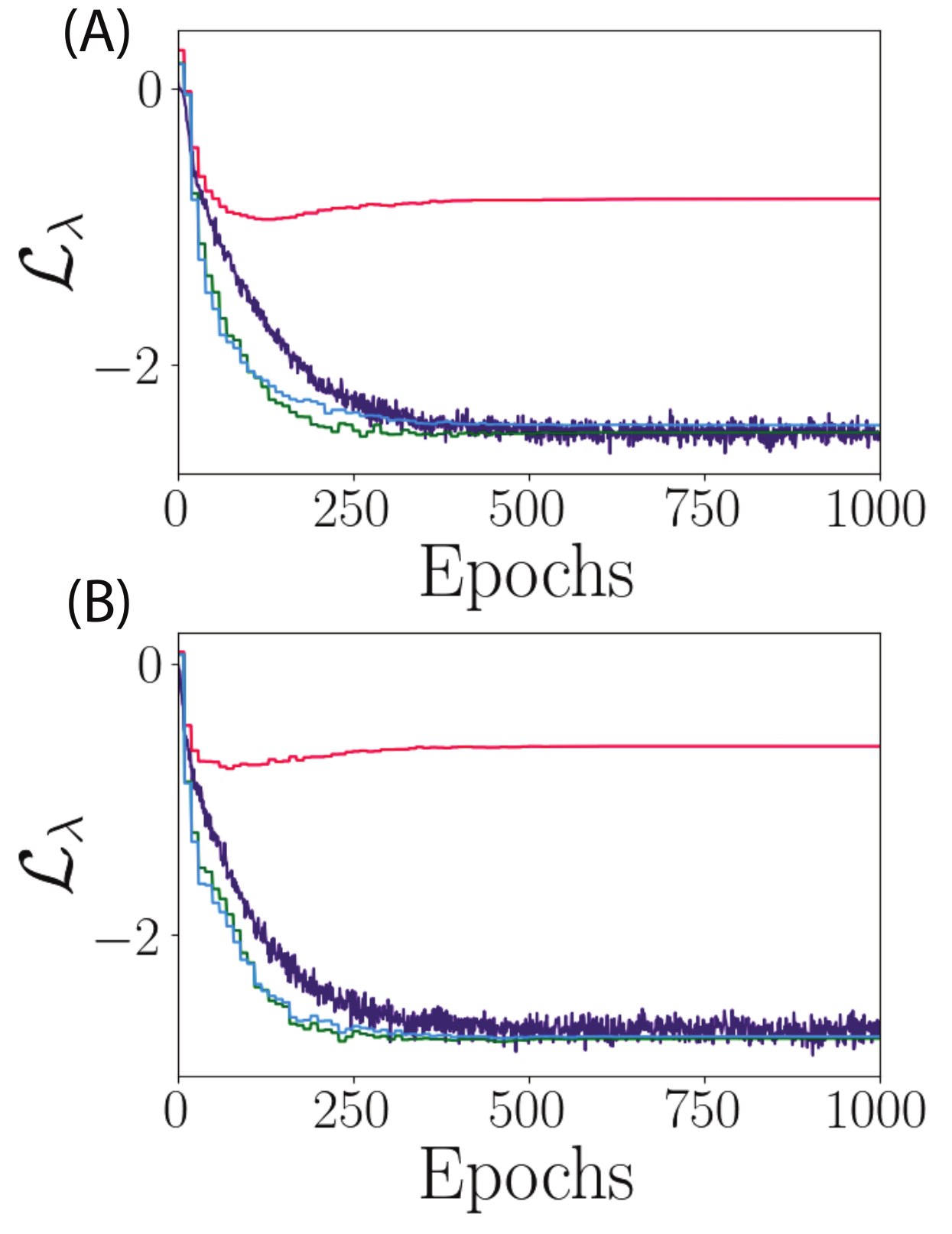}
    \caption{Optimization curves for the time-dependent committor for the unfolding (A) and folding (B) reaction, for a model trained on the 147 cartesian descriptors. The red curve corresponds to the 1st cumulant of the validation set, the purple curve corresponds to the 1st cumulant of the test set, and the green and blue curves correspond to the exponential and 2nd cumulant estimate of the test set. }
    \label{fig:cartesian_loss}
\end{figure}

For the first set of optimizations, we used the bare coordinates due to the lack of available descriptors and the anisotropy of the system. A neural network comprising of 2 layers of 50 units each with a CELU activation function connected to a final layer with a sigmoid activation function for the final layer was used for representing the time-dependent committor. A dropout layer with a frequency of $0.2$ was used for regularization. The time was represented through the function $f(t) = 1/(t_f - t + 0.01)$ based on the limiting form of the exact controller. For this specific case, we used a scheduler to tune the learning rate based on the validation loss to prevent overfitting. Optimization was performed using the RMSProp optimizer for a 1000 steps. Due to the small gap in the spectrum of the generator, and the poor choice of representation, convergence of the loss function was found to be extremely challenging with a limited set of samples as visible from the loss curves for the optimization for the unfolding and folding case in Fig. \ref{fig:cartesian_loss}. The 1st cumulant estimate for the for the time-lagged training set is found to be $2.5$ and $2.7$ for the unfolding and the folding reaction, and the validation loss is found to be $0.9$ and $0.8$ for the unfolding and folding reaction. The second cumulant approximation to Eq.~\ref{eq:rate_OM} and its direct evaluation, are computed for the validation set is identical to the training loss function. All of these estimators are considerably off from the empirical value of $\ln kt_f$ that is found to be $-4.8 \pm 0.3$ and $-5.9 \pm 0.3$ for the unfolding and folding cases respectively.

It is possible to get a considerably better model by tuning the ansatz, decreasing the time-lag, or adding more samples.
Since our goal is to simply to get the leading descriptors and to perform a dynamically consistent coarse-graining, rather than solving the full body Backward Kolmogorov equation, we perform the optimization on the first $5$ normal modes that is discussed within the main text, to confirm whether the reactive modes that were predicted by the cartesian optimizations indeed are accurate enough to perform optimal control.

\begin{figure}[!ht]
  \centering
    \includegraphics[width=0.49\textwidth]{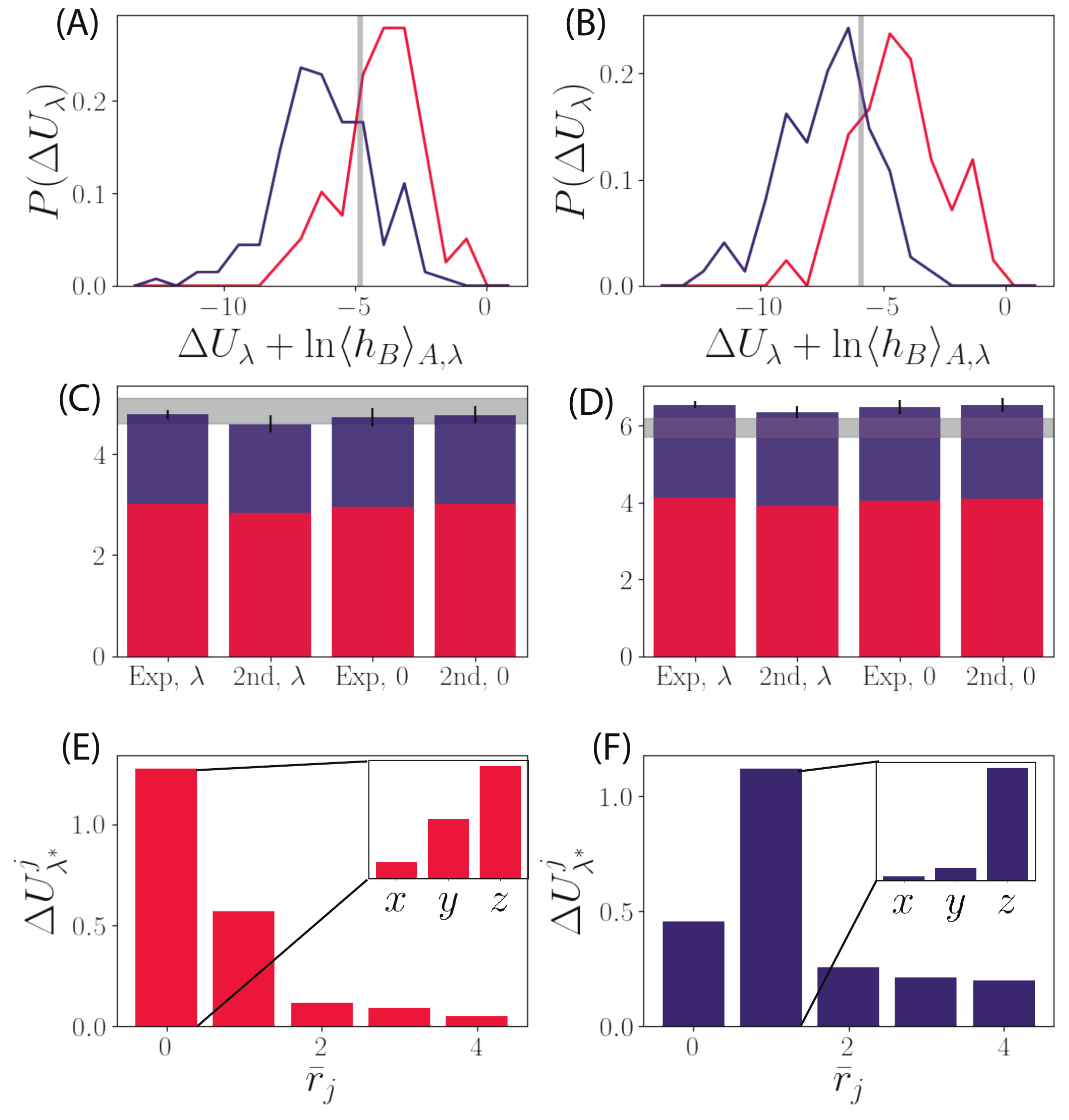}
    \caption{Validation of the controller for the nonequilibrium folding and unfolding reaction that is trained on the first 5 normal modes. Shifted relative action distribution within the driven reactive (purple) and the original reactive ensembles (red) for the unfolding reaction (A) and folding (B) reaction. The shift is given by the log reactivity in the driven ensemble. The four estimators based on the 2nd cumulant expansion and the exponential average of the OM action averaged within the driven reactive ($\lambda$) and the original (0) reactive trajectory ensemble for the unfolding (C) and folding reaction (D). The red bars denote the value of the estimator and the purple bars denote the additional correction given by $\ln \langle h_B \rangle_{A,\lambda}$. The gray lines correspond to the empirical estimate of $\ln kt_f$ obtained from brute force trajectories. Normal mode decomposition of the optimized action averaged within the conditioned ensemble for the unfolding (E) and folding (F) reaction. }
    \label{fig:poly_validate}
\end{figure}

For training, 16 descriptors including the 15 modes and the descriptor for time $f(t)$ are passed into the same architecture as the Cartesian optimization. For this case, we did not observe any overfitting and we performed optimization over 20000 steps with a cosine annealing scheduler~\cite{loshchilov2016sgdr}. Post optimization, we evolved 1000 trajectories using the optimized controller for validation. The reactivity is found to be $\approx$ 20\% for the unfolding case and $\approx$ 10\% for the folding case, which while still not close to optimal is still higher than the original systems by 2 orders of magnitude. We also found that this reactivity can be tuned by increasing the conditioning time. To validate the controller, we considered the action distribution for the driven and the undriven process is shown in Fig. \ref{fig:poly_validate} (A) and (B) respectively. Both of them show a strong overlap, and intersect close to the empirical value of $\ln kt_f$, suggesting that they are accurate. The accuracy is further verified by considering the second and exponential estimators within the driven and original reactive trajectory ensembles. For the estimate, only 50 undriven reactive trajectories and between 100-200 driven reactive trajectories are used for both the cases based on the reactivity of the driven ensembles. All the estimators are shown in Fig. \ref{fig:poly_validate} (C) and (D) for the unfolding and folding cases, with errorbars denoting standard errors computed from 5 sets of trajectories randomly selected from the ensemble. All estimators are found to be accurate compared against the empirical estimate of $\ln kt_f$. 

Finally, we consider the decomposition of the action along the normal modes and visualize it in Fig. \ref{fig:poly_validate} (E) and (F) for the unfolding and folding reactions. The decompositions are found in excellent agreement with the results from the model trained on the Cartesian modes. The leading descriptors are found to be same, validating the dominant modes that are activated for the reactions. While the coupling of other nonlinear modes to the transition is possible, including importance of the momentum degrees of freedom that are not necessarily guaranteed to homogenize for nonequilibrium steady-states, these results validate the use of this method for performing dynamically consistent coarse graining of systems far from equilibrium. 

\bibliography{NoneqReactive}

\end{document}